\documentclass[12pt]{iopart}

\usepackage{graphicx}
\usepackage{color}
\usepackage{mathrsfs}
\usepackage{bm}
\begin{document}
\title{Deriving the trajectory equations of gyrocenter with a multi-parameter Lie transform method}

\author{Shuangxi Zhang}

\address{
 Graduate School of Energy Science, Kyoto University, Uji, Kyoto 611-0011, Japan. }
\ead{zshuangxi@gmail.com;zhang.shuangxi.3s@kyoto-u.ac.jp}

\date{\today}

\begin{abstract}
It's pointed out that the values of the generators derived by the modern gyrokinetic theory are inappropriately amplified by the pullback transform with the existence of electromagnetic perturbation, and the trajectory equations of the gyrocenter include terms containing the perturbative magnetic vector potential not in the form of its curl. These terms  are not physical ones, as the perturbative magnetic vector potential could include an arbitrary gauge term. In this paper, instead of the single-parameter Lie transform method utilized by the modern gyrokinetic theory, a multi-parameter Lie transform method is utilized. All characteristic scales of the system can be taken into account. With the assistance of a kind of linear cancellation rule, the new method can decouple the gyroangle $\theta$ from the remaining degrees of freedom up to the second order of the parameter as the ratio between the Larmor radius and the transverse length scale of the equilibrium magnetic field. With this new method, those nonphysical terms don't appear anymore in the trajectory equations.
\end{abstract}


\maketitle

\section{Introduction}\label{sec1}

Modern Gyrokinetic theory (GT) is a strong theoretical tool to help the numerical simulation of the magnetized plasma based on the following  facts: modern GT reduces the six-dimensional phase space of distribution function to a five-dimensional one with the magnetic moment as a constant for each particle \cite{1983cary,1988hahm,1990brizard}; the time step can be larger than that for the full-orbit simulation for the low-frequency electrostatic perturbation\cite{1983wwlee, wwleejcp1987,1998zhihonglin,2010garbet,2008qin1,idomuranf2009}.

Modern GT utilizes the Lie transform perturbation theory (LTPT) to decouple the gyroangle $\theta$, which is a fast variable, from the remaining degrees of freedom\cite{1988hahm,1990brizard,2007brizard1,2009cary,1982littlejohn,1983littlejohn,2000qinhong1,1999honqingpop,2014parra,2011parra, 2010scott}. During the decoupling procedure, the other scales of perturbation take roles, as shown in Sec.(\ref{sec6}).
A strongly magnetized plasma system includes the following several kinds of perturbation\cite{1980antonsen,1982frieman, 2009fujisawa1, 2016chenliu, 2011ido, 2005diamond1}. The first one is  the gyrating motion, which is characterized by the parameter $\varepsilon\equiv \rho/L_0$. $\rho$ is the Larmor radius and $L_0$ is the characteristic length scale of the equilibrium magnetic field. The second one is the amplitude of each component of the fluctuating magnetic field $\mathbf{B}_w$ or $\nabla  \times {{\bf{A}}_w}$, where $\mathbf{A}_w$ is the perturbation of the magnetic potential. The amplitude of $\mathbf{A}_w$ can not be treated as a parameter, since $\mathbf{A}_w$ includes an arbitrary gauge term. The perturbation $\mathbf{B}_w$ can be magnetic island, Alfven waves or other electromagnetic perturbation. The third one is the amplitude of each component of the fluctuating electric field $\mathbf{E}$ or $\nabla \phi$, where $\phi$ is the perturbation of the electric potential. The fourth one is the spatial gradient length of $\mathbf{A}_w$ and $\phi$ in each direction. The fifth one is the oscillating frequency of the perturbation.
In strongly magnetized plasma, the amplitude of the normalized $\mathbf{B}_w$ and $\bf{E}$ are small quantities and can be treated as perturbations \cite{2012xu,2009fujisawa1,2011ido,2011sharapov,2013kongdefeng,2011bottino}. The length scale of the spatial gradient of the fluctuation can be much smaller than that of the equilibrium magnetic field and even of scale of the Larmor radius. Meanwhile, the oscillating frequency of the perturbation, especially the external incident high frequency wave can be of a frequency equal to even higher than the gyrofrequency. Both spatial scale and frequency scale of the perturbation take part in the procedure of decoupling $\theta$ angle from the remaining degrees of freedom.

The sing-parameter LTPT adopted by modern GT is first given in Ref.(\cite{1983cary}). As can be observed in Sec.(\ref{app1}), one character of this single-parameter LTPT is that it  includes only one parameter. But as mentioned before, charged particles included in the magnetized plasma experience perturbations of several characteristic scales. Two classical papers which utilize the single-parameter LTPT to solve the perturbation experienced by charged particles in the strong magnetic field are Ref.(\cite{1988hahm}) and (\cite{1990brizard}).
Ref.(\cite{1988hahm}) considers  only the electrostatic fluctuation, the amplitude of which is restricted to be of order $O(\varepsilon)$, where $\varepsilon$ is the value of normalized Larmor radius. However, the amplitude of the normalized $\phi$ in fact usually deviates from $O(\varepsilon)$ as experiment and numerical simulation show. It should be treated as an independent perturbation parameter. Therefore, Ref.(\cite{1988hahm}) only considers a special case. Ref.(\cite{1990brizard}) considers electrostatic and electromagnetic perturbation. The amplitude of both is the same and treated as an independent perturbation parameter different from $\varepsilon$. Since this method is mainly adopted by the following researchers to solving the electromagnetic and electrostatic perturbation \cite{2010scott,2000qinhong1,2009krommes,2009miyato}, our analysis focuss on this method, which is called  modern GT in this paper.

The two parameters modern GT facing include $\varepsilon$ and the amplitude of $\mathbf{A}_w$ and $\phi$. The latter is recorded as $\varepsilon_w$ in this paper.  Since the single-parameter LTPT has only one parameter, modern GT solves the two parameters $\varepsilon$ and $\varepsilon_w$ separately. Firstly, it's assumed that the system only includes a equilibrium magnetic field configuration, so that only $\varepsilon$ appears in the system. By imposing the single-parameter LTPT over the Lagrangian 1-form of this system to cancel $\theta$-coupling term up to $\varepsilon^2$, the guiding center Lagrangian is derived. Next, other perturbation characterized by the parameter $\varepsilon_w$ is introduced to the guiding center Lagrangian. By treating $\varepsilon_w$ as the perturbation parameter (ordering parameter), whilst  $\varepsilon$ as a normal quantity, the gyrocenter Lagrangian is derived. However, such a procedure leads to the order confusion which will be pointed out in the context. The consequence resulted from the order confusion is that after carrying out the coordinate transform, the order of generators is not consistent with the prescribed order. Specifically, the order of the generators are amplified by $1/\varepsilon$ times which causes the coordinate transform not to be a near identical transform (NIT).

Another issue modern GT faces is that the equations of motin for gyrocenter include terms like $\nabla \mathbf{A}_w$. Such kind of terms are nonphysical, since $\mathbf{A}_w$ includes an arbitrary gauge term  as $\nabla f$ with $f$ being an arbitrary function of gyrocenter coordinate. The term $\nabla^2 f$ generated by $\nabla \mathbf{A}_w$ as a nonphysical term can't be cancelled. The correct form for the appearance of $\mathbf{A}_w$ should be like $\nabla \times \mathbf{A}_w$.

In this paper, we utilize a new multi-parameter Lie transform method to decouple the $\theta$ angle from the remaining degrees of freedom when the perturbations exist. This method is first given by Ref.(\cite{2016shuangxi4}). It assumes that the new 1-form can be formulated by a formal formula like $\Gamma \left( {{\bf{Z}},{\bf{E}}} \right) = \left[ {\exp \left( { - {\bf{E}}\cdot{L_{\bf{g}}}} \right)\gamma } \right]\left( {\bf{Z}} \right)$. $\bf{E}$ is a parameter vector defined as ${\bf{E}} \equiv ({\varepsilon _1},{\varepsilon _2},{\varepsilon _3} \cdots )$. $\mathbf{g}$ is the vector of generators defined as ${\bf{g}} \equiv \left( {{\mathbf{g}_1},{\mathbf{g}_2},{\mathbf{g}_3}, \cdots } \right)$. The inner product is defined to be ${\bf{E}}\cdot{L_{\mathbf{g}}} \equiv \sum\limits_{n = 1} {{\varepsilon _i}{L_{{\mathbf{g}_i}}}}$. The action of $L_{\mathbf{g}_i}$ over 1-form $\gamma$ is defined as ${ {{L_{{\mathbf{g}_i}}}\gamma } } $ and will be given in Sec.(\ref{sec0}). This analytical formula assisted by a kind of linear cancellation rule can decouple the $\theta$ angle from the remaining degrees of freedom by cancelling out those $\theta$-coupling terms order by order, without introducing the order confusion. During the cancellation procedure, $\bf{E}$ and $\bf{g}$ are resolved, without any inappropriate amplification of $\mathbf{g}_i$. Eventually, it's found that $\theta$ can be decoupled from the remaining degrees of freedom up to $\varepsilon^{2}$. Besides, the trajectory equations don't include nonphysical term like $\nabla \mathbf{A}_w$ and new  terms of the effect of spatial gradient and frequency of the perturbations are derived.

The rest of the paper is arranged as follows. In Sec.(\ref{sec0}), a simple introduction to some rules of Lie derivative and pullback transform of differential forms is given. In Sec.(\ref{sec2}), we give a derivation of gyrocenter kinetics based on Modern GT. In Sec.(\ref{sec3}), some comments about the results derived in Sec.(\ref{sec2}) are given. In Sec.(\ref{sec4}), a new Lie transform method given in Ref.(\cite{2016shuangxi4}) is chosen based on the features of the Lagrangian differential 1-form of the charged particle chosen from magnetized plasma. Sec.(\ref{sec5}) is the preparation work before carrying out the multi-parameter Lie transform method.
In Sec.(\ref{sec6}), the multi-parameter Lie transform method is applied to the Lagrangian 1-form to decouple gyroangle from the remaining degrees of freedom. The coordinate transform is solved in Sec.(\ref{sec8}). Sec.(\ref{sec7}) is  discussion and summary.

\section{A simple introduction to some rules of Lie derivative and pullback transform of differential form}\label{sec0}

This paper will utilize several rules of Lie derivative and pullback transform of differential forms, which are introduced as follows.
For a vector field $G = \left( {{g^1},{g^2}, \cdots ,{g^p}} \right)$  with the basis ${{\bf{e}}_i} = \frac{\partial }{{\partial {y^i}}}, i\in \{1,\cdots,p\}$, and the differential 1-form $\gamma=\sum\limits_{k = 1}^p {{\gamma _k}\left( {\bf{y}} \right)d{y^k}}$ defined on the manifold $\mathbf{y}$ with its coordinate frame also written as $\bf{y}$, the Lie derivative for single ${{\bf{g}}^i} = {g^i}\frac{\partial }{{\partial {y^i}}}$ acting upon $\gamma$ is
\begin{equation}\label{vp7}
{L_{{{\bf{g}}^i}}}\left( {\sum\limits_{k = 1}^p {{\gamma _k}\left( {\bf{y}} \right)d{y^k}} } \right) = \sum\limits_{k = 1}^p {\left[ \begin{array}{l}
{g^i}\left( {{\partial _i}{\gamma _k} - {\partial _k}{\gamma _i}} \right)d{y^k}\\
 + \frac{\partial }{{\partial {y^k}}}\left( {{g^i}{\gamma _i}} \right)d{y^k}
\end{array} \right]} ,
\end{equation}
where repeated indexes don't denote summation. $\gamma_k$ is called the $k$th component of $\gamma$ throughout this paper. It should be noticed that the Lie derivative over differential 1-form in Eq.(\ref{vp7}) generates a full differential term as the last term on R.H.S of Eq.(\ref{vp7}). This term doesn't contribute to the dynamics in Lagrangian. This property will be repeatedly used in the context.
It's easy to check the following linear property of the Lie derivative of 1-form
\begin{equation}\label{vp9}
{L_{{{\bf{g}}^i} + {{\bf{g}}^j}}}\left( {{\gamma _k}\left( {\bf{y}} \right)d{y^k}} \right) = \left( {{L_{{{\bf{g}}^i}}} + {L_{{{\bf{g}}^j}}}} \right)\left( {{\gamma _k}\left( {\bf{y}} \right)d{y^k}} \right).
\end{equation}
The Lie derivative of external or full differential term is
\begin{equation}\label{vp11}
{L_{{{\bf{g}}^i}}}\left( {d\gamma } \right) = d\left( {{L_{{{\bf{g}}^i}}}\gamma } \right),
\end{equation}
which means the Lie derivative of full differential term is still a full differential term. This property will be repeatedly applied in the context.

As for the pullback transform, it's first assumed that $\phi_{\varepsilon}$ is a coordinate transformation defined as ${\phi_\varepsilon }:{\bf{Y}} \to {\bf{y}}$ , both of which are $n$-dimensional manifolds. $\varepsilon$ is a small parameter. $\gamma \left( {{\bf{y}},\varepsilon } \right) = \sum\limits_{h \ge 0} {{\varepsilon ^h}} \sum\limits_{k = 1}^p {{\gamma _{hk}}\left( {\bf{y}} \right)} d{y^k}$ is a differential 1-form defined on $\bf{y}$, where the subscripts $h$,$k$ denote the order and the component of the 1-form, respectively. $\mathbf{v}_j$ is a tangent vector defined on $\mathbf{Y}$ with $j\in \{1,2,\cdots,p\}$. Here, $\mathbf{v}_j$ is chosen to be $\mathbf{v}_j=\partial/\partial Y_j$. $\phi$ induces a pullback transformation ${\phi^*_\varepsilon}$ of 1-form $\gamma$ written as $\Gamma=\phi^*_\varepsilon \gamma$ defined on $\mathbf{Y}$.  With the standard mathematical terminology, the pullback transformation is defined as
\begin{equation}\label{a1}
{\left. {\left( {{\phi_\varepsilon ^*}\gamma(\mathbf{y},\varepsilon) } \right)} \right|_{\bf{Y}}}\left( {\frac{\partial }{{\partial {Y^j}}}} \right) = {\left. \gamma(\mathbf{y},\varepsilon)  \right|_{\bf{y}}}\left( {{{\left. {d\phi_\varepsilon \left( {\frac{\partial }{{\partial {Y^j}}}} \right)} \right|}_{\bf{y}}}} \right),
\end{equation}
based on the contraction rule between the tangent vector and the cotangent vector \cite{2006marsden, 1989arnoldbook}. $\phi^*$ satisfies the linear property as
\begin{equation}\label{vp30}
{\phi_\varepsilon ^*}\left( {{\gamma _{hi}}d{y^i} + {\gamma _{lj}}d{y^j}} \right) = {\phi_\varepsilon ^*}\left( {{\gamma _{hi}}d{y^i}} \right) + {\phi_\varepsilon ^*}\left( {{\gamma _{lj}}d{y^j}} \right).
\end{equation}
The formula ${d\phi_\varepsilon \left( {\frac{\partial }{{\partial {Y^j}}}} \right)}$ is a pushforward transformation of $\partial/\partial Y^j$ and equals $d\phi_\varepsilon \left( {\frac{\partial }{{\partial {Y^j}}}} \right) = \frac{{\partial \phi _{\varepsilon}^k\left( {\bf{Y}} \right)}}{{\partial {Y^j}}}\frac{\partial }{{\partial {y^k}}}$. Substituting the pushforward transformation to Eq.(\ref{a1}) and applying the contraction rule, the $j$th component of the new 1-form transformed from ${\varepsilon ^h}\sum\limits_{k = 1}^p {{\gamma _{hk}}\left( {\bf{y}} \right)}dy^k$  is
\begin{equation}\label{a2}
{\varepsilon ^h}{\Gamma _{hj}}({\bf{Y}},\varepsilon ) = {\varepsilon ^h}\sum\limits_{k = 1}^p {{\gamma _{hk}}({\phi _\varepsilon }({\bf{Y}}) )\frac{{\partial \phi _\varepsilon ^k\left( {\bf{Y}} \right)}}{{\partial {Y^j}}}}.
\end{equation}

\section{Simple introduction to deriving the equations of motion of the gyrocenter based on Modern GT}\label{sec2}

The purpose modern GT pursues is to find a NIT with the gyroangle $\theta$ in the new coordinate system decoupled from the remaining degrees of freedom up to some order of the small parameter characterizing the amplitude of the perturbations. Modern GT uses the single-parameter LTPT\cite{1983cary} to realize this purpose\cite{1988hahm, 1990brizard}. The mathematical procedure of Modern GT to derive the new Lagrangian on gyrocenter coordinate is as follows. The detailed derivation can also be found in Ref.(\cite{1990brizard}).

\subsection{Normalizing physical quantities}\label{sec2.1}

Modern GT first imposes the single-parameter over the Lagrangian 1-form
\begin{equation}\label{a119}
\gamma ' = \left( {q{\bf{A}}\left( {{{\bf{x}}}} \right) + m{\bf{v}}} \right)\cdot d{\bf{x}} - \frac{1}{2}m{v^2}dt
\end{equation}
which determines the kinetics of the charged particle chosen from the magnetized plasma.
$(\mathbf{x},\mathbf{v})$ is the real physical coordinate frame.
By decoupling gyroangle $\theta$ from the remaining degrees of freedom up to $O(\varepsilon^2)$ with $\varepsilon  = \frac{\rho }{{{L_0}}}$, the guiding center Lagrangian 1-form can be derived as
\begin{eqnarray}\label{g88}
{\gamma _0} = && \left( {q{\bf{A}}\left( {{{\bf{X}}_{\bf{1}}}} \right) + m{U_1}{\bf{b}}} \right)\cdot d{{\bf{X}}_1} + \frac{m}{q}{\mu _1}d{\theta _1} \nonumber \\
&&- ({\mu _1}B\left( {{{{\bf{X}}}_{{1}}}} \right) + \frac{1}{2}m U_1^2)dt.
\end{eqnarray}
Then, the differential 1-form for the perturbation wave is introduced
\begin{equation}\label{g89}
{\gamma _w} = q{{\bf{A}}_w}({{\bf{X}}_1} + {\bm{\rho }},t)\cdot d\left( {{{\bf{X}}_1} + {\bm{\rho }}} \right) - q\phi_w \left( {{{\bf{X}}_1} + {\bm{\rho }},t} \right)dt,
\end{equation}
with ${\bm{\rho }} = {{\bm{\rho }}'_0} + ( \cdots )$ and ${\bm{\rho}' _0} = \frac{1}{q}\sqrt {\frac{{2m{\mu _1}}}{{B\left( {{{\bf{X}}_1}} \right)}}} \left( { - {{\bf{e}}_1}\cos \theta  + {{\bf{e}}_2}\sin \theta } \right)$. $(\mathbf{b},\mathbf{e}_1,\mathbf{e}_2)$ form a right-hand cartesian coordinate frame are will be ignored. Symbol $'(\cdots)'$ means higher order terms. $\mathbf{A}_w,\phi_w$ are the symbols for  perturbations of  magnetic potential and electric potential, respectively. $\mathbf{A}(\mathbf{X_1})$ is the equilibrium magnetic potential.  The guiding center coordinates is denoted as ${{\bf{Z}}}_1 = ({{\bf{X}}}_1,{U}_1,{\mu}_1 ,{\theta}_1)$. The other notations in Eqs.(\ref{g88},\ref{g89}) can be found in Ref.(\cite{1990brizard}).

The test particle is chosen from a thermal equilibrium plasma ensemble, e.g., the thermal equilibrium plasma in tokamak. Therefore, $\mathbf{A},U_1,\mathbf{X}_1,t,\mathbf{B}, \phi_w, \mu$ can be normalized by $A_0=B_0 L_0,v_t,L_0,L_0/v_t,B_0,A_0 v_t, mv_t^2/B_0$, respectively. $B_0, L_0$ are the characteristic amplitude and spatial length of magnetic field, respectively. $v_t$ is the thermal velocity of the particle ensemble containing the test particle. To conform to the procedure modern GT utilizes, it's assumed that the amplitude of $\mathbf{A}_w$ takes the role of the small parameter. The small parameter representing the normalized amplitude of $\mathbf{A}_w,\phi_w$ is extracted out, so that $\mathbf{A}_w,\phi_w$ is reformulated to be $\varepsilon_w \mathbf{A}_w, \varepsilon_w \phi_w$, respectively, with $O(|\mathbf{A}_w|)\sim O(|\phi_w|)\sim O(1)$. Throughout the rest of the paper, all physical quantities are normalized.

With $\mathbf{A}_w,\phi_w$ changed to be $\varepsilon_w \mathbf{A}_w, \varepsilon_w \phi_w$, the normalized editions of  Eqs.(\ref{g88}) and (\ref{g89}) are
\begin{eqnarray}\label{g2}
{\gamma _0} =&& {\bf{A}}\left( {{{\bf{X}}_1}} \right)\cdot d{{\bf{X}}_1} + \varepsilon {U_1}{\bf{b}}\cdot d{{\bf{X}}_1} + {\varepsilon ^2}{\mu _1}d{\theta _1} \nonumber \\
&& - \varepsilon \left( {\frac{{U_1^2}}{2} + {\mu _1}B\left( {{{\bf{X}}_1}} \right)} \right)dt,
\end{eqnarray}
and
\begin{eqnarray}\label{g3}
&&{\gamma _w} = \varepsilon_w {{\bf{A}}_w}\left( {{{\bf{X}}_1} + \bm{\rho} ,t} \right)\cdot d\left( {{{\bf{X}}_{\bf{1}}} + \bm{\rho} } \right) + \varepsilon_w {\phi _w}\left( {{{\bf{X}}_{\bf{1}}} + \bm{\rho} ,t} \right)dt \nonumber \\
&& \approx \varepsilon_w \exp \left( {{\varepsilon \bm{\rho} _0}\cdot\nabla } \right){{\bf{A}}_w}\left( {{{\bf{X}}_1},t} \right)\cdot\left( \begin{array}{l}
d{{\bf{X}}_1} + \frac{{\varepsilon\partial { \bm{\rho} _0}}}{{\partial {{\bf{X}}_1}}}\cdot d{{\bf{X}}_1} \nonumber \\
 + \frac{{\varepsilon \partial {\bm{\rho} _0}}}{{\partial {\mu _1}}}d{\mu _1} + \frac{{\varepsilon \partial { \bm{\rho} _0}}}{{\partial {\theta _1}}}d{\theta _1}
\end{array} \right)  \nonumber \\
&&  - \varepsilon_w \exp \left( {{\varepsilon \bm{\rho} _0}\cdot\nabla } \right){\phi _w}\left( {{{\bf{X}}_1},t} \right)dt,
\end{eqnarray}
respectively, where ${\varepsilon } \equiv \frac{{m{v_t}}}{{{A_0}q}}=\frac{{{\rho _t}}}{{{L_0}}},{\bm{\rho} _t} = \frac{{m{v_t}}}{{{B_0}q}},{\bm{\rho} _0} = \sqrt {\frac{{2\mu }}{{B\left( {\bf{X}}_1 \right)}}} \left( { - {{\bf{e}}_1}\cos \theta  + {{\bf{e}}_2}\sin \theta } \right)$. The normalized edition of $\rho$ is $\varepsilon \rho_0$ with $O(\rho_0)=O(1)$. In Eqs.(\ref{g2},\ref{g3}), both $\varepsilon,\varepsilon_w$ take part in the calculation. If the small parameters $\varepsilon,\varepsilon_w$ are just used as a symbol for the order of terms, they will be written as $\varepsilon^*, \varepsilon_w^*$ with a superscript $*$ for comparison. This rule is adopted throughout the rest of this paper.

\subsection{Carrying out the pullback transform and deriving the  trajectory equations of the gyrocenter}\label{sec2.2}

$\gamma_{0}+\gamma_{w}$ is the total Lagrangian differential 1-form with $\rho_0$ depending on the fast angle $\theta$.  To decouple $\theta$ from the remaining degrees of freedom, the single-parameter LTPT given BY \ref{app1} is adopted with $\varepsilon_w$ treated as the small parameter, while $\varepsilon$ as a normal quantity not involved in the order expanding. The gyrocenter frame is recorded as ${\bf{Z}} = \left( {{\bf{X}},\mu ,U,\theta } \right)$.
The coordinate transform should satisfy NIT and is formally recorded  as ${{\bf{Z}}_1} = \exp \left( { - {\varepsilon _w}{g^i}\left( {\bf{Z}} \right){\partial _i}} \right){\bf{Z}}$ with  $O(g^i)\sim O(1)$ for all $i\in (\mathbf{X},U,\mu,\theta)$. All of $g^i$s need to be solved. The new $\Gamma$ induced by this coordinate transform is
\begin{equation}\label{a99}
\Gamma=[\cdots T_2 T_1(\gamma_0+\gamma_w)](\mathbf{Z})+dS,
\end{equation}
with ${T_i} = \exp \left( { - \varepsilon _w^j{L_{{{\bf{g}}_j}}}} \right)$f. $\mathbf{g}_j$ includes elements $g_j^i$ for $i\in \{\mathbf{X},\mu,U,\theta\}$.
By expanding $\Gamma$ in Eq.(\ref{a99}) as the sum like
\begin{equation}
\Gamma  = \sum\limits_{n\ge 0} {\frac{1}{{n!}}\varepsilon _w^n{\Gamma _n}},
\end{equation}
$O(1)$ part of the new $\Gamma$ is
\begin{equation}\label{g35}
{\Gamma _0} = {{\bf{A}}}\left( {\bf{X}} \right)\cdot d{\bf{X}} + {\varepsilon }U{\bf{b}}\cdot d{\bf{X}} + {\varepsilon ^{2}}\mu d\theta
 - {\varepsilon }H_0 dt,
\end{equation}
with $H_0={\frac{{U_{}^2}}{2} + \mu B\left( {\bf{X}} \right)}$. The $O(\varepsilon_w)$ part is
\begin{eqnarray}\label{g36}
{\varepsilon _w}{\Gamma _1} =&& \left( { - \left( {{\bf{B}} + \varepsilon U\nabla  \times {\bf{b}}} \right) \times \left( {{\varepsilon _w}{{\bf{g}}^X}} \right) - \varepsilon {\varepsilon _w}{g^U}{\bf{b}} + \exp \left( {{\varepsilon \bm{\rho} _0}\cdot{\nabla _{\bf{X}}}} \right)\left( {{\varepsilon _w}{{\bf{A}}_w}} \right)} \right)\cdot d{\bf{X}} \nonumber \\
&& + \varepsilon \left( {{\varepsilon _w}{{\bf{g}}^X}\cdot{\bf{b}}} \right)dU + \left( {\exp \left( {{\varepsilon \bm{\rho} _0}\cdot{\nabla _{\bf{X}}}} \right)\left( {{\varepsilon _w}{{\bf{A}}_w}} \right)\cdot\frac{{\partial {\varepsilon \bm{\rho} _0}}}{{\partial \theta }} - {\varepsilon ^2}{g^\mu }} \right)d\theta  \nonumber \\
&& - \left( {{\varepsilon ^2}\left( {{\varepsilon _w}{g^\theta }} \right) + \exp \left( {\varepsilon {\bm{\rho} _0}\cdot{\nabla _{\bf{X}}}} \right)\left( {{\varepsilon _w}{{\bf{A}}_w}} \right)\cdot\frac{{\varepsilon\partial { \bm{\rho} _0}}}{{\partial \mu }}} \right)d\mu  \nonumber \\
&& - \left( { - \left( {{\varepsilon _w}{{\bf{g}}^X}\cdot\nabla {H_0}} \right) - \varepsilon {\varepsilon _w}U{g^U} - \varepsilon {\varepsilon _w}{g^\mu }B + \exp \left( {{\varepsilon \bm{\rho} _0}\cdot{\nabla _{\bf{X}}}} \right)\left( {{\varepsilon _w}{\phi _w}} \right)} \right)dt  \nonumber \\
&& + {\varepsilon _w}dS.
\end{eqnarray}
Eq.(\ref{g36}) obviously shows the confusion between the order of $\varepsilon$ and $\varepsilon_w$.

Modern GT requires the following identities
\begin{equation}\label{g34}
\Gamma_{1i}=0,i\in \{\mathbf{X},U,\mu,\theta\}.
\end{equation}
All of $g^i$s can be derived based on Eqs.(\ref{g36},\ref{g34}) as
\begin{eqnarray}\label{g37}
{{\bf{g}}^X} = && - \frac{1}{{{\bf{b}}\cdot{{\bf{B}}^*}}}\left( {{\bf{b}} \times \exp \left( {\varepsilon \bm{\rho}_0 \cdot\nabla } \right){{\bf{A}}_w}\left( {{\bf{X}},t} \right)  + {\bf{b}} \times \nabla {S_1}} \right)  \nonumber \\
&&- \frac{{{{\bf{B}}^*}}}{{{\varepsilon }}}\frac{{\partial {S_1}}}{{\partial U}}
\end{eqnarray}
where ${{\bf{B}}^*} = {\bf{B}} + {\varepsilon }U\nabla  \times {\bf{b}}$,
\begin{equation}\label{g38}
{g^U} = \frac{1}{{{\varepsilon }}}{\bf{b}}\cdot\exp \left( {\varepsilon \bm{\rho}_0 \cdot\nabla } \right){{\bf{A}}_w}\left( {{\bf{X}},t} \right) + \frac{1}{{{\varepsilon }}}{\bf{b}}\cdot\nabla {S_1},
\end{equation}
\begin{equation}\label{g39}
{g^\mu } = \frac{1}{{{\varepsilon}}}\exp \left( {\varepsilon \bm{\rho}_0 \cdot\nabla } \right){{\bf{A}}_w}\left( {{\bf{X}},t} \right)\cdot\frac{{ \partial \bm{\rho}_0 }}{{\partial \theta }} + \frac{1}{{{\varepsilon ^{2}}}}\frac{{\partial {S_1}}}{{\partial \theta }},
\end{equation}
\begin{equation}\label{g40}
{g^\theta } =  - \frac{1}{{{\varepsilon}}}\exp \left( {\varepsilon \bm{\rho}_0 \cdot\nabla } \right){{\bf{A}}_w}\left( {{\bf{X}},t} \right)\cdot\frac{{\partial \bm{\rho}_0 }}{{\partial \mu }} - \frac{1}{{{\varepsilon ^{2}}}}\frac{{\partial {S_1}}}{{\partial \mu }}.
\end{equation}
\begin{eqnarray}\label{g41}
\frac{{\partial {S_1}}}{{\partial t}} + U{\bf{b}}\cdot\nabla {S_1} + \frac{1}{{{\varepsilon }}}\frac{{\partial {S_1}}}{{\partial \theta }} = F + {\Gamma _{1t}},
\end{eqnarray}
where
\begin{eqnarray}\label{g92}
F = && \exp \left( {\varepsilon \bm{\rho}_0 \cdot\nabla } \right){\phi _w}\left( {{\bf{X}},t} \right) \nonumber \\
&& -  U{\bf{b}}\cdot\exp \left( {\varepsilon \bm{\rho}_0 \cdot\nabla } \right){{\bf{A}}_w}\left( {{\bf{X}},t} \right) \nonumber \\
&&- {B(\mathbf{X})}\exp \left( {\varepsilon \bm{\rho}_0 \cdot\nabla } \right){{\bf{A}}_w}\left( {{\bf{X}},t} \right)\cdot\frac{{\partial \bm{\rho}_0 }}{{\partial \theta }}.
\end{eqnarray}
The smaller term ${{{\bf{g}}^X}\cdot \varepsilon \nabla {H_0}}$ and other higher order terms are ignored in Eq.(\ref{g41}).

For the low frequency perturbation, inequalities $\left| {\frac{{\partial {S_1}}}{{\partial t}}} \right| \ll \left| {\frac{B}{{{\varepsilon }}}\frac{{\partial {S_1}}}{{\partial \theta }}} \right|, \left| {U{\bf{b}}\cdot\nabla {S_1}} \right| \ll \left| {\frac{B}{{{\varepsilon }}}\frac{{\partial {S_1}}}{{\partial \theta }}} \right|$ hold, and the lowest order equation of Eq.(\ref{g41}) is
\begin{eqnarray}\label{g42}
\frac{B(\mathbf{X})}{{{\varepsilon }}}\frac{{\partial {S_1}}}{{\partial \theta }} = F +\Gamma_{1t}.
\end{eqnarray}
To avoid the secularity of $S_1$ over the integration of $\theta$, $\Gamma_{1t}$ is chosen to be
\begin{equation}\label{g43}
{\Gamma_{1t}} = -\left\langle F \right\rangle,
\end{equation}
where $\left \langle F \right \rangle $ means the averaging over $\theta$.
The new $\Gamma$ approximated up to $O(\varepsilon_w)$ is
\begin{equation}\label{a100}
\Gamma  = \left( {{\bf{A}}\left( {\bf{X}} \right) + \varepsilon U{\bf{b}}} \right)\cdot d{\bf{X}} + {\varepsilon ^2}\mu d\theta  - \left( {{H_0} - {\varepsilon _w}{\Gamma _{1t}}} \right)dt,
\end{equation}
with ${H_0} = \varepsilon \left( {\frac{{{U^2}}}{2} + \mu B\left( {\bf{X}} \right)} \right)$ and ${{\Gamma _{1t}}}$ given in Eq.(\ref{g43}).

By imposing the minimal action principle over the action as the integral of the Lagrangian 1-form given by Eq.(\ref{a100}) over the time, the equations of motion can be derived as
\begin{equation}\label{a115}
\mathop {\bf{X}}\limits^. {\rm{ = }}\frac{{U{{\bf{B}}^*} + {\bf{b}} \times \nabla \left( {{H_0} - {\varepsilon _w}{\Gamma _{1t}}} \right)}}{{{\bf{b}}\cdot{{\bf{B}}^*}}},
\end{equation}
\begin{equation}\label{a116}
\dot U = \frac{{ - {{\bf{B}}^*}\cdot\nabla \left( {{H_0} - {\varepsilon _w}{\Gamma _{1t}}} \right)}}{{{\varepsilon}{\bf{b}}\cdot{{\bf{B}}^*}}}.
\end{equation}

\section{Comments on the results given in Sec.(\ref{sec2})}\label{sec3}

\subsection{Violation of NIT by the coordinate transform given in Sec.(\ref{sec2})}\label{sec3.1}

\subsubsection{Inappropriate amplification of the generators}\label{sec3.1.1}

Now we check that whether the coordinate transform given in Sec.(\ref{sec2}) is a NIT. In other words, whether $O(g^i)\sim O(1)$ holds for $i\in \{\mathbf{X}, \mu, U,\theta\}$. For convenience, only the pure perturbative electromagnetic potential is considered, so that the electric field only includes the inductive part and no electrostatic part exists.

To get the order sequence of $g^i$s, we first derive the order sequence of $S_1$, the equation of which is given in Eq.(\ref{g42})
with $\Gamma_{1t}$ given in Eq.(\ref{g43}). The order sequence of $S_1$ is
\begin{equation}\label{g63}
\varepsilon_w {S_1} = \varepsilon_w{\varepsilon ^{* 2}}( \cdots ) + \varepsilon_w{\varepsilon ^{*3 }}( \cdots ) +  \cdots.
\end{equation}
The superscript $*$ of ${\varepsilon ^{*2}}$  represents the order of $(\cdots)$ adjacent to it as explained before.

The lowest order term of all of $g^i$s should be of the order equaling or higher than $O(1)$ to satisfy NIT. Substituting the order sequence of $S_1$ into Eqs.(\ref{g38})-(\ref{g41}), the order sequence of $g^i$s can be derived. The lowest order of $\mathbf{g}^{X}$ is $O(1)$, which is produced by the lowest order term of the exponential expansion of the first term on the right of Eq.(\ref{g37}). The lowest order of $g^U$, $g^{\mu}$ and $g^{\theta}$ is $O(\varepsilon^{-1})$ and also originates from the lowest order term of the exponential expansion of the first term on the right of Eq.(\ref{g38}),(\ref{g39}) and (\ref{g40}), respectively. The coordinate transform for $U, \mu, \theta$ are approximately as $U_1\approx U-\varepsilon_w g_1^U, \mu_1 \approx \mu -\varepsilon_w g_1^\mu, \theta_1\approx \theta-\varepsilon_w g_1^\theta$. It's observed that for a perturbation with the amplitude being $O(\varepsilon_\omega)$, the coordinate transform amplifies the generators by $1/\varepsilon$ times to get the new coordinate. This coordinate transform doesn't satisfy NIT, as $\varepsilon$ is a very small quantity.

One consequence of the violation of NIT is as follows. In numerical and theoretical applications, the following transform between the distribution functions in the full-orbit coordinate and the gyrocenter coordinate is frequently applied
\begin{eqnarray}
f\left( {{{\bf{x}}},{\mu _1},{u_1},t} \right) = \int \begin{array}{l}
F\left( {{\bf{X}},\mu ,U,t} \right)\delta \left( {{{\bf{x}}} - {\bf{X}} - \varepsilon{\bm{\rho }_0}} \right)\\
 \times \delta \left( {\mu  - {\mu _1}} \right)\delta \left( {U - {u_1}} \right)B(\mathbf{X}){d^3}{\bf{X}}d\mu dUd\theta, \nonumber.
\end{array}
\end{eqnarray}
However, it's noticed that ${\varepsilon _w}g_1^U,{\varepsilon _w}g_1^\mu ,{\varepsilon _w}g_1^\theta$ given by Eqs.(\ref{g38}-\ref{g40}) are of order $O(\varepsilon_w/\varepsilon)$. So the integrand of the this integral transform should take the following formula
\begin{eqnarray}
&& \delta \left( {{{\bf{x}}} - {\bf{X}} - \varepsilon {{\bm{\rho }}_0}} \right)\delta \left( {\mu  - {\varepsilon _w}g_1^\mu  - {\mu _1}} \right) \nonumber \\
&& \times \delta \left( {U - {\varepsilon _w}g_1^U - {u_1}} \right)\delta \left( {\theta  - {\varepsilon _w}g_1^\theta  - {\theta _1}} \right). \nonumber
\end{eqnarray}
To make modern GT self-consistent, we need to remove the violation of NIT from the coordinate transform.

%
%

\subsection{The equations of motion not consistent with the real physics}\label{sec3.2}

In Eqs.(\ref{a115}) and (\ref{a116}), the contribution of the perturbation are mainly $\frac{{{\varepsilon _w}{\bf{b}} \times \nabla \left\langle {U{\bf{b}}\cdot\exp \left( {\varepsilon {\rho _0}\cdot\nabla } \right){{\bf{A}}_w}\left( {{\bf{X}},t} \right)} \right\rangle }}{{{\bf{b}}\cdot{{\bf{B}}^*}}}$ and $\frac{{{\varepsilon _w}{{\bf{B}}^*}\cdot\nabla \left\langle {U{\bf{b}}\cdot\exp \left( {\varepsilon {\rho _0}\cdot\nabla } \right){{\bf{A}}_w}\left( {{\bf{X}},t} \right)} \right\rangle }}{{\varepsilon {\bf{b}}\cdot{{\bf{B}}^*}}}$, respectively. Both are not physical terms, since $\mathbf{A}_w$ includes an arbitrary gauge term like $\nabla f(x)$. The gradient operator in both terms can not cancel the gauge term.  The real physical contribution should be like $\frac{{\partial {{\bf{A}}_{\bf{w}}}/\partial t \times {\bf{b}}}}{{{\bf{b}}\cdot{B^*}}}$ and $- {\bf{b}}\cdot\frac{\partial }{{\partial t}}{{\bf{A}}_w}$, which are the $\mathbf{E}\times \mathbf{B}$ drift produced by inductive electric field, and parallel inductive electric field acceleration of  the charged particle. Therefore, Eqs.(\ref{a115},\ref{a116}) need to be modified.


\section{Choosing the appropriate LTPT}\label{sec4}

\subsection{The restriction imposed on LTPT by the characters of the Lagrangian 1-form}\label{sec4.1}

Based on the discussion in the previous context, there are two main characters of the Lagrangian 1-form. \\

(1). The perturbations present multiple scales. The scales include the amplitude of $\nabla \times \mathbf{A}_w, \partial \mathbf{A}_w /\partial t$, and $\nabla \phi $, the spatial gradient scale and the oscillating frequency of the perturbation wave.

(2). Since $\nabla \times \mathbf{A}_w$ is the physical quantity, the trajectory equations of gyrocenter should  include  terms like $\nabla \times \mathbf{A}_w$ as well as terms as action of operators over $\nabla \times \mathbf{A}_w$.\\

The two characters limit the transform scheme of LTPT. Point (1) requires a multi-parameter transform. Point (1) also requires the expansion of new 1-form should be done order by order for each parameter to avoid inappropriate amplification of those generators, as explained in Sec.(\ref{sec3.1.1}).


\subsection{Utilizing a new multi-parameter Lie transform method in Ref.(\cite{2016shuangxi4})}\label{sec4.2}

Based on the discussion in Sec.(\ref{sec4.1}), we utilize a new LTPT, the detail of which is given in Ref.(\cite{2016shuangxi4}). Here, only a simple introduction is given. Firstly, the 1-form in the old coordinate frame is written as $\gamma(\mathbf{z},\mathbf{E}_n)$ with $\mathbf{E}_n=\{\varepsilon_1,\cdots,\varepsilon_n\}$. $\{\varepsilon_1,\cdots,\varepsilon_n\}$ are the basic parameter set and the element in it is independent  from each other, e.g, the set of parameter $(\varepsilon_1,\varepsilon_1\varepsilon_2,\varepsilon_2,\varepsilon_2^2)$ only includes basic parameter set $(\varepsilon_1,\varepsilon_2)$. Then, $\gamma(\mathbf{z},\mathbf{E}_n)$ can be expanded formally like
\begin{equation}\label{af24}
\gamma ({\bf{z}},\mathbf{E}_n) = \sum\limits_{{m_1} \ge 0, \cdots ,{m_n} \ge 0} {\varepsilon _1^{{m_1}}\varepsilon _2^{{m_2}} \cdots \varepsilon _n^{{m_n}}{\gamma _{{m_1}{m_2} \cdots {m_n}}}} ({\bf{z}}).
\end{equation}
All of perturbation parameters are already extracted out from $\gamma_{m_1 m_2 \cdots m_n}$. So $O(\gamma_{m_1 m_2 \cdots m_n})\sim O(1)$ holds.


This method assumes a formally analytical formula of the new 1-form transformed from $\gamma(\mathbf{z},\mathbf{E}_n)$
\begin{equation}\label{af26}
\Gamma(\mathbf{Z},\mathbf{E}_n)  = \sum\limits_{{m_1} \ge 0, \cdots ,{m_n} \ge 0} {\varepsilon _1^{{m_1}}\varepsilon _2^{{m_2}} \cdots \varepsilon _n^{{m_n}}{\Gamma _{{m_1}{m_2} \cdots {m_n}}}}(\mathbf{Z},\mathbf{E}_n),
\end{equation}
with
\begin{equation}\label{af25}
{\Gamma _{{m_1}{m_2} \cdots {m_n}}}\left( {{\bf{Z}},{\bf{E}}_n} \right) = [\exp \left( { - {\bf{E}}\cdot{L_{\bf{g}}}} \right){\gamma _{{m_1}{m_2} \cdots {m_n}}}]({\bf{Z}}) + d{S_{{m_1}{m_2} \cdots {m_n}}}({\bf{Z}}).
\end{equation}
Here, $\bf{E}$ is a formal parameter vector and $\bf{g}$ is a formal generator vector. The elements in $\mathbf{E}$ are combinations of the elements in $\mathbf{E}_n$. The inner product ${{\bf{E}}\cdot{L_{\bf{g}}}}$ is defined to be the summation of all ${E_i}{L_{{\mathbf{g}_i}}}$ where $E_i$ and $\mathbf{g}_i$ are the $i$th elements of $\bf{E}$ and $\bf{g}$.
Each element of $\mathbf{g}$ is of order $O(1)$.
$\bf{E}$ and $\bf{g}$ are solved based on the following procedure.

First, it needs to expand the exponential formula $[\exp \left( { - {\bf{E}}\cdot{L_{\bf{g}}}} \right)\gamma](\bf{Z}) $ order by order. For each order, those terms depending on fast variables need to be cancelled. To cancel these terms at some order, $\epsilon=\varepsilon_1^{l_1}\cdots \varepsilon_n^{l_n}$ and $\mathbf{g}_{l_1\cdots l_n}$ with $l_1\ge 0,\cdots,l_n \ge 0$ are introduced into exponential formula to get $\exp \left( { - {\bf{M}} - \varepsilon _1^{{l_1}} \cdots \varepsilon _n^{{l_n}}{L_{{{\bf{g}}_{{l_1} \cdots {l_n}}}}}} \right)$ where $\bf{M}$ denotes the already existing operators. The newly introduced generators generate  new terms depending on fast variables, the lowest order terms of which are designed to cancel the already existed terms depending on fast variables at the same order. Therefore, the left terms of fast variables are of higher order than that of those cancelled. By repeatedly introducing new generators for cancellation, the left $\theta$-coupling terms in the new 1-form are of higher and higher order, and $\varepsilon_1^{l_1}\cdots \varepsilon_n^{l_n}$s and $\mathbf{g}_{l_1\cdots l_n}$s are solved order by order by the way.

For specific Lagrangian 1-form, there exists some rule to help introducing those $\epsilon=\varepsilon_1^{l_1}\cdots \varepsilon_n^{l_n}$s and $\mathbf{g}_{l_1\cdots l_n}$s to realize the cancellation purpose. Ref.(\cite{2016shuangxi4}) introduces a linear cancellation rule for the Lagrangian 1-form of motion of charged particle immersed in strong magnetic field.


After solving $\bf{E}$ and $\bf{g}$, the coordinate transform can be derived based on the pullback transform of Eq.(\ref{a2}) with $\varepsilon$ replaced by $\mathbf{E}_n$, as explained in Ref.(\cite{2016shuangxi4}).

\section{Preparation before applying the multi-parameter LTPT to deriving trajectory equations of gyrocenter}\label{sec5}

\subsection{Normalizing Lagrangian 1-form}\label{sec5.1}


The Lagrangian 1-form for a test particle in the real physical coordinate system is
\begin{eqnarray}\label{a10}
\gamma  =&& \left( {{{\bf{A}}}\left( {\bf{x}} \right) + {{\bf{A}}_\alpha }\left( {{\bf{x}},t} \right)} \right)\cdot d{\bf{x}} + {\varepsilon }{\bf{v}}\cdot d{\bf{x}} \nonumber \\
&& - {\varepsilon }\frac{{{{\bf{v}}^2}}}{2}dt - {\phi _\sigma }\left( {{\bf{x}},t} \right)dt,
\end{eqnarray}
where  the symbols for the perturbation of magnetic vector potential and electric potential are changed to be $\mathbf{A}_{\alpha}$ and $\mathbf{\phi}_\sigma$, respectively. The meanings of $\alpha$ and $\sigma$ are given later.  In Eq.(\ref{a10}), all quantities are already normalized by the scheme given in Sec.(\ref{sec2}). The prerequisites of this normalization is that the test particle is chosen from a thermal equilibrium plasma ensemble, and $v_t$ is the thermal velocity of the ensemble. $v_t$ can also be taken place of by a average velocity of the particle during a period time if the test particle is just a single particle.

\subsection{The ordering for various scales of the perturbation}

In Eq.(\ref{a10}), $\mathbf{A}, \mathbf{A}_\alpha,\mathbf{\phi}_\sigma$ represent equilibrium magnetic potential, the perturbation of magnetic vector potential and the perturbation of the electric potential, respectively. However, they are only symbols and their amplitude can't be used as the ordering parameter.
It will be shown in our new theory that only $\nabla \times \mathbf{A}_{\alpha}, \partial \mathbf{A}_{\alpha}/\partial t, \nabla \phi $ contribute to the trajectory equations. Here, the power indexes over $\varepsilon$ are used to denote the order of each independent scale:  $O(\nabla  \times {{\bf{A}}_\alpha }) =O({\varepsilon ^\alpha })$, denoting the order of the amplitude of the perturbation magnetic field; $O(\nabla {\phi _\sigma }) = O({\varepsilon ^\sigma })$, denoting the order of the amplitude of the electrostatic field; $O(\partial {{\bf{A}}_\alpha }/\partial t) =O({\varepsilon ^\eta })$, denoting the amplitude of inductive electric field; $O({\partial _{{X^i}}}\ln |{{\bf{A}}_\alpha }|) = O({\partial _{{X^i}}}\ln |{\phi _\sigma }|) = O({\varepsilon ^{ - \beta }})$ for $i\in \{1,2,3\}$, denoting the order of the spatial scale of the perturbation; $O({\partial _t}\ln |{{\bf{A}}_\alpha }|) = O({\partial _t}\ln |{\phi _\sigma }|) = O({\varepsilon ^{ - \tau }})$, denoting the order of the frequency scale of the perturbations. And $O\left( {\frac{{{\partial _{(\mu ,U,\theta)}}F({\bf{Z}})}}{{F({\bf{Z}})}}} \right) = O(1)$ with $F(\mathbf{Z})$ being an any function of $\bf{Z}$ adopted in the rest of this paper. $(\alpha ,\beta ,\sigma ,\tau ,\eta ) > 0$ hold and $\alpha ,\beta ,\sigma ,\tau ,\eta$ are independent from each other.

In this paper, it's assumed that the spatial gradient of perturbation over each $\mathbf{e}_i$ direction for $i\in (1,2,3)$ is the same, and $B_{\alpha}^i=(\nabla \times \mathbf{A}_{\alpha})^i$ is of the same amplitude for $i\in (1,2,3)$. For the real physical systems, e.g., tokamak plasma, the spatial gradient of the perturbation and the amplitude of $B_{\alpha}^i$ is different for different directions, respectively. For such kind of a problem, the following method can also be applied.

\subsection{Transform of velocity coordinate}\label{sec5.2}

The first operation is to transform $(\mathbf{x},\mathbf{v})$ to $(\mathbf{x},u_1,\mu_1,\theta)$, where $u_1$ is parallel velocity and $\mu_1$ is magnetic moment. $u_1=\mathbf{v}\cdot \mathbf{b}$ and $\mu_1=v_\perp^2/2B(\mathbf{x})$, and ${\widehat {\bf{v}}_ \bot } = \left( {{{\bf{e}}_1}\sin \theta  + {{\bf{e}}_2}\cos \theta } \right)$. $(\mathbf{e}_1,\mathbf{e}_2,\mathbf{b})$ are orthogonal mutually. This transformation just transforms the velocity vector from rectangular coordinate system to cylindrical coordinate system. After this transformation, $\gamma$ becomes
\begin{eqnarray}\label{a11}
\gamma  = \left( \begin{array}{l}
{{\bf{A}}}\left( {\bf{x}} \right) +{{\bf{A}}_{\alpha}}\left( {{\bf{x}},t} \right) + {\varepsilon}{u_1}{\bf{b}}  \nonumber \\
 + {\varepsilon }\sqrt {2B({\bf{x}}){\mu _1}} {\widehat {\bf{v}}_ \bot }
\end{array} \right)\cdot d{\bf{x}}  \nonumber \\
 - {\varepsilon }\left( {\frac{{u_1^2}}{2} + {\mu _1}B({\bf{x}})} \right)dt - {\phi _\sigma }\left( {{\bf{x}},t} \right)dt,
\end{eqnarray}
which can be separated into four parts
\begin{equation}\label{a12}
{\gamma _0} = {{\bf{A}}}\left( {\bf{x}} \right)\cdot d{\bf{x}},
\end{equation}
\begin{eqnarray}\label{a13}
{\gamma _1} = \left( {u_1{\bf{b}} +\sqrt {2B({\bf{x}})\mu_1 } {{\widehat {\bf{v}}}_ \bot }} \right)\cdot d{\bf{x}} - \left( {\frac{{{u_1^2}}}{2} + \mu_1 B({\bf{x}})} \right)dt,
\end{eqnarray}
\begin{equation}\label{a14}
{\gamma _{0\sigma} } =  - \phi_\sigma \left( {\bf{x}},t \right)dt,
\end{equation}
\begin{equation}\label{b9}
{\gamma _{0\alpha} } = {{\bf{A}}_{\alpha}}\left( {{\bf{x}},t} \right)\cdot d{\bf{x}}.
\end{equation}

\subsection{Linear cancellation rule}\label{sec.lin}

As discussed in Sec.(\ref{sec4.2}), the complex cancellation calculation may be summarized as some simple rules. For Lagrangian system of charged particle moving in strong magnetic field, the linear cancellation rule is given in Ref.(\cite{2016shuangxi4}), which is repeated here.

Non-$\theta$-coupling terms $\Upsilon  = {\bf{A}}\cdot d{\bf{X}} + {\Gamma _{1{\bf{X}}\parallel }}\cdot d{\bf{X}} + {\Gamma _{1t}}dt$ with ${\Gamma _{1{\bf{X}}\parallel }} = \varepsilon U{\bf{b}},{\Gamma _{1t}} = \varepsilon (\frac{{{U^2}}}{2} + \mu B({\bf{X}}))$ are eventually kept and independent from any generator. There terms are determined by the properties of the physical systems, e.g., the charge and mass per particle; the amplitude and configuration of equilibrium magnetic field. Other terms like  electromagnetic and electrostatic fluctuation change with time. There terms form the cornerstone of linear cancellation rule which is given as follows.

(1). Assuming that $[\Gamma_{o\chi \mathbf{X}}]_{\perp} \cdot d\mathbf{X}$ is a 1-form only including $\bf{X}$ component and  $[\Gamma_{o\chi \mathbf{X}}]_\perp$ is a $\theta$-coupling term  perpendicular to the unit vector of magnetic field $\bf{b}$, and its order is $O(\varepsilon^\chi)$ as the subscript $n$ indicates, we introduce a generator field  $\mathbf{g}_{\chi}^{\bf{X}}$ perpendicular to $\mathbf{b}$, to the exponential operator to form $\exp ( - {\varepsilon ^{\chi }}{L_{\mathbf{g}_{\chi}^{\mathbf{X}}}} +  \cdots )\Upsilon$, which generates a linear term ${\varepsilon ^\chi}{\bf{g}}_\chi^{\bf{X}} \times {\bf{B}}\left( {\bf{X}} \right)\cdot d{\bf{X}}$. This term is the lowest order term among all the generated terms and is used to cancel $[\Gamma_{o\chi \mathbf{X}}]_\perp$.


(2). Assuming that $[\Gamma_{o\chi \mathbf{X}}]_{\parallel} \cdot d\mathbf{X}$ is a differential 1-form and  $[\Gamma_{o\chi \mathbf{X}}]_\parallel$ is a $\theta$-coupling term  parallel to $\bf{b}$, and its order is $O(\varepsilon^\chi)$ as the subscript $\chi$ indicates, we introduce a generator  $g_{\chi-1}^U$ to the exponential operator to form $\exp ( - {\varepsilon ^{\chi - 1}}{L_{g_{\chi - 1}^U}} +  \cdots )\Upsilon$, by which the lowest-order terms generated is a linear term set
\begin{equation}\label{af88}
\begin{array}{l}
{\bf{N}}_{\chi}^U =  - {\varepsilon ^{\chi - 1}}{L_{g_{\chi - 1}^U}}\Upsilon  \\
 =  - {\varepsilon ^\chi}g_{\chi - 1}^U{\partial _U}{\Gamma _{1{\bf{X}}\parallel }}\cdot d{\bf{X}} - {\varepsilon ^n}g_{\chi - 1}^U{\partial _U}{\Gamma _{1t}}dt+d S,
\end{array}
\end{equation}
The first term in ${\bf{N}}_{\chi}^U$ is used to cancel $[\Gamma_{o\chi,\mathbf{X}}]_\parallel$.

(3). Assuming that $\Gamma_{o\chi,t}dt $  is a differential 1-form only including $t$ component and $\Gamma_{o\chi t}$ is a $\theta$-coupling term and its order is $\varepsilon^\chi$, we introduce a generator $g^{\mu}_{\chi-1}$ to the exponential operator to form $\exp ( - {\varepsilon ^{\chi - 1}}{L_{g_{\chi - 1}^\mu }} +  \cdots )\Upsilon$, by which the lowest-order term generated is a linear term
\begin{equation}\label{af89}
\begin{array}{l}
{\bf{N}}_{\chi }^\mu  =  - {\varepsilon ^{\chi - 1}}{L_{g_{\chi - 1}^\mu }}\Upsilon \\
 =  - {\varepsilon ^\chi}g_{\chi - 1}^\mu {\partial _\mu }{\Gamma _{1t}}dt+dS,
\end{array}
\end{equation}
which is used to cancel $\Gamma_{o\chi t}$.

Based on points $(1),(2),(3)$, all linear terms in $\mathbf{N}_{\chi}^{U},\mathbf{N}_{\chi}^{\mu}$ are completely cancelled. The left $\theta$-coupling terms in $\mathbf{N}_{\chi}^{\mathbf{X}}$ are of order higher than $O(\varepsilon^{\chi})$. By repeatedly carrying out the linear cancellation, the order of left $\theta$-coupling terms becomes higher and higher.

%

\section{Applying multi-parameter LTPT to derive trajectory equations of gyrocenter }\label{sec6}

In this section, multi-parameter LTPT is adopted to deriving the dynamic equations of charged particle immersed in strong magnetic field with electromagnetic perturbation with gyroangle $\theta$ decoupled from the remaining degrees of freedom up to some order.

\subsection{Cancelling out $\theta$-coupling terms of $O(\varepsilon)$}\label{sec5.3.1}

For this method, the formula of new 1-form is assumed to be $\Gamma  = \left[ {\exp \left( { - {\bf{E}}\cdot{L_{\bf{g}}}} \right)\gamma } \right]\left( {\bf{Z}} \right) + dS\left( {\bf{Z}} \right)$. Then, $\Gamma$ is expanded based on the order of each parameters. To calculate the Lie derivative during the expanding, the following rules will be adopted frequently
\begin{eqnarray}\label{m1}
{L_{{\bf{g}}_1^{\bf{x}}}}\left( {{\bf{f}}({\bf{Z}})\cdot d{\bf{X}}} \right) =  - {\bf{g}}_1^{\bf{x}} \times \nabla  \times {\bf{f}}\left( {\bf{Z}} \right)\cdot d{\bf{X}} \nonumber \\
 - {\bf{g}}_1^{\bf{x}}\cdot\left( {{\partial _t}{\bf{f}}({\bf{Z}})dt + {\partial _\theta }{\bf{f}}({\bf{Z}})d\theta  + {\partial _\mu }{\bf{f}}({\bf{Z}})d\mu } \right) + dS,
\end{eqnarray}
and
\begin{eqnarray}\label{m2}
{L_{{\bf{g}}_1^{\bf{x}}}}\left( {h({\bf{Z}})dt} \right) = {\bf{g}}_1^{\bf{x}}\cdot\nabla h\left( {\bf{Z}} \right)dt+dS, \nonumber
\end{eqnarray}
where $\mathbf{f}(\bf{Z})$ and $h(\bf{Z})$ are arbitrary vector and scalar function of $\bf{Z}$, respectively.

The first three terms are
\begin{equation}\label{a70}
{\Gamma _0} = {\gamma _0}(\mathbf{Z}),
\end{equation}
\begin{equation}\label{a72}
{\Gamma _{0\sigma} } =\gamma_{0\sigma}(\mathbf{Z})=  - {\phi _\sigma }\left( {{\bf{X}},t} \right)dt,
\end{equation}
\begin{equation}\label{a73}
{\Gamma _{0\alpha} } = \gamma_{0\alpha}(\mathbf{Z})={{\bf{A}}_\alpha }\left( {{\bf{X}},t} \right)\cdot d{\bf{X}}.
\end{equation}
The $O(\varepsilon)$ part of new $\Gamma$ is recorded as
\begin{equation}\label{a71}
{\Gamma _1}(\mathbf{Z}) = d{S_1}(\mathbf{Z}) + {\gamma _1}(\mathbf{Z}) + {\left( ? \right)_1},
\end{equation}
where order parameter $\varepsilon$  is divided out in both sides. Such an operation will be adopted later. ${\left( ? \right)_1}$ means terms of $O(\varepsilon)$ generated by newly introduced generators.
$\hat{\mathbf{v}}_\perp$ includes gyroangle $\theta$. The $\theta$-coupling term  included in $\gamma_1$ is $\sqrt {2\mu B({\bf{X}})} {\widehat {\bf{v}}_ \bot }$ which belongs to  $\bf{X}$ component and is perpendicular to $\mathbf{b}$. Based on the linear cancellation rule, to cancel this term, it needs to introduce a generator $\mathbf{g}_1^{\mathbf{X}}$.

The introduction of $\mathbf{g}_1^\mathbf{X}$ generates new terms. The terms independent on $\mathbf{g}_1^{\mathbf{X}}$ are recorded to be
\begin{eqnarray}\label{a17}
{{\bf{M}}_0} =&& \left( {{\bf{A}}\left( {\bf{X}} \right) + {{\bf{A}}_\alpha }\left( {{\bf{X}},t} \right)+{\varepsilon }U{\bf{b}} + {\varepsilon }\sqrt {2B({\bf{X}})\mu } {{\widehat {\bf{v}}}_ \bot }} \right)\cdot d{\bf{X}}  \nonumber \\
&& - {\varepsilon }\left( {\frac{{{U^2}}}{2} + \mu B({\bf{X}}) + {\phi _\sigma }({\bf{X}},t)} \right)dt.
\end{eqnarray}
The terms linear to $\mathbf{g}_1^{\mathbf{X}}$ are recorded as
\begin{eqnarray}\label{a18}
\begin{array}{l}
{\bf{M}}_1^{\bf{X}} =  - \varepsilon {L_{{\bf{g}}_1^{\bf{X}}}}\gamma \\
 = \varepsilon {\bf{g}}_1^{\bf{X}} \times {\bf{B}}\left( {\bf{X}} \right)\cdot d{\bf{X}} + {\varepsilon ^2}{\bf{g}}_1^{\bf{X}} \times \left( {\nabla  \times {\gamma _{1{\bf{X}}}}} \right)\cdot d{\bf{X}}\\
 + {\varepsilon ^2}{\bf{g}}_1^{\bf{X}}\cdot{\partial _\mu }{\gamma _{1{\bf{X}} \bot }}d\mu  + {\varepsilon ^2}{\bf{g}}_1^{\bf{X}}\cdot{\partial _\theta }{\gamma _{1{\bf{X}} \bot }}d\theta  - {\varepsilon ^2}{\bf{g}}_1^{\bf{X}}\cdot\nabla {\gamma _{1t}}dt\\
 + {\varepsilon ^{*(1 + \alpha )}}\left( {\varepsilon {\bf{g}}_1^{\bf{X}} \times {{\bf{B}}_\alpha }\left( {\bf{X}} \right)} \right)\cdot d{\bf{X}} + {\varepsilon ^{*\left( {1 + \eta } \right)}}\left( {\varepsilon {\bf{g}}_1^{\bf{X}}\cdot{\partial _t}{\gamma _{0\alpha {\bf{X}}}}} \right)dt\\
 - {\varepsilon ^{*\left( {1 + \sigma } \right)}}\left( {\varepsilon {\bf{g}}_1^{\bf{X}}\cdot\nabla {\gamma _{0\sigma t}}} \right)dt.
\end{array}
\end{eqnarray}
Since equations ${\bf{g}}_1^{\bf{X}}\cdot{\partial _U}{\gamma _{1{{\bf{X}}_\parallel }}}=0$ hold, there is no $U$ component in ${\bf{M}}_1^{\bf{X}}$.


The introducing of $\mathbf{g}_1^\mathbf{X}$ makes $(?)_1$ equal to  ${L_{{\mathbf{g}_1^{\mathbf{X}}}}}{\gamma _0}$  which is the first term in $\mathbf{M}_1^{\mathbf{X}}$. The    $\bf{X}$ component of $\Gamma_1$ becomes
\begin{eqnarray}\label{a19}
{\Gamma _{1{\bf{X}}}} =\nabla S_1 + {{\bf{g}}_1} \times {{\bf{B}}}+  {U{\bf{b}} + \sqrt {2\mu B({\bf{X}})} {{\widehat {\bf{v}}}_ \bot }},
\end{eqnarray}
To cancel the $\theta$-coupling term in Eq.(\ref{a19}), it's derived that
\begin{equation}\label{a22}
{\mathbf{g}_1^{{\bf{X}}}} = -\bm{\rho}
\end{equation}
where $\bm{\rho} = \sqrt {\frac{{{2\mu }}}{{B\left( {{{\bf{X}}}} \right)}}} \left( { - {{\bf{e}}_1}\cos \theta  + {{\bf{e}}_2}\sin \theta } \right)$ and
\begin{equation}\label{a23}
{\Gamma _{1}} =dS_1+ U\cdot d{\bf{X}} -  \left( {\frac{{{U^2}}}{2} + \mu B\left( {\bf{X}} \right)} \right)dt.
\end{equation}

Then, we need to cancel the $\theta$-coupling terms of order higher than $O(\varepsilon)$ generated by $\mathbf{g}_1^{\mathbf{X}}$.


\subsection{Cancelling out $\theta$-coupling terms of order $O(\varepsilon^{1+\sigma})$ in new $\Gamma$}\label{sec5.3.2}

The term of $O(\varepsilon^{1+\sigma})$ in $\mathbf{M}_1^{\mathbf{X}}$ can be reformulated as $ - \mathbf{g}_1^{\mathbf{X}}\cdot \nabla {\gamma _{0\sigma t}}d{t} = {\bf{g}}_1^{\bf{X}}\cdot\nabla {\phi _\sigma }d{t}$. $-\nabla \phi_\sigma$ is a physical term representing electrostatic electric field and its order is $O(\varepsilon^\sigma)$  given previously. ${\bf{g}}_1^{\bf{X}}\cdot\nabla {\phi _\sigma }d{t}$ is a $\theta$-coupling term.  To cancel this term, it needs to introduce a generator $g^\mu_{\sigma}$ to get $\exp \left( { - \varepsilon {L_{{\bf{g}}_1^{\bf{X}}}} - {\varepsilon ^\sigma }{L_{g_\sigma ^\mu }}} \right)\gamma \left( {\bf{Z}} \right)$, by which the lowest-order terms generated are $\mathbf{N}_{1+\sigma}^\mu$ and a non-linear term $\frac{1}{2}{L_{g_\sigma ^\mu }}{L_{{\bf{g}}_1^{\bf{X}}}}{\gamma _0}$.  The sum of this two terms are recorded as
\begin{equation}\label{a54}
\begin{array}{l}
{\bf{M}}_{ \sigma }^\mu  =  - {\varepsilon ^{1 + \sigma }}{L_{g_\sigma ^\mu }}{\Gamma _1} - \frac{1}{2}{\varepsilon ^{1 + \sigma }}g_\sigma ^\mu {\partial _\mu }{\gamma _{1{{\bf{X}}_ \bot }}}\cdot d{\bf{X}}\\
 =  - {\varepsilon ^{1 + \sigma }}g_\sigma ^\mu {\partial _\mu }{\gamma _{1t}}dt - \frac{1}{2}{\varepsilon ^{1 + \sigma }}g_\sigma ^\mu {\partial _\mu }{\gamma _{1{{\bf{X}}_ \bot }}}\cdot d{\bf{X}}
\end{array}
\end{equation}
where ${\gamma _{1{\bf{X}} \bot }} \equiv \sqrt {2B({\bf{X}})\mu } {\widehat {\bf{v}}_ \bot }$ and $L_{\mathbf{g}_1^{\mathbf{X}}}(\gamma_0)(\mathbf{X})= {\bf{g}}_1^{\bf{X}} \times {\bf{B}} \cdot d\mathbf{X}=  - {\gamma _{1{{\bf{X}}_ \bot }}}\cdot d\bf{X}$ is applied.
The first term in ${\bf{M}}_\sigma ^\mu $ is used to cancel ${\bf{g}}_1^{\bf{X}}\cdot\nabla {\phi _\sigma }d{t}$. The left term in ${\bf{M}}_\sigma ^\mu $ can be canceled by introducing generator $\mathbf{g}_{1+\sigma}^{\bf{X}}$. The introduction of $\mathbf{g}_{1+\sigma}^{\bf{X}}$ generates a set of linear term recorded as $\mathbf{M}_{1+\sigma}^{\bf{X}}$ similar to $\mathbf{M}_1$. The first term of $\mathbf{M}_{1+\sigma}^{\bf{X}}$ is used to cancel  $- \frac{1}{2}g_\sigma ^\mu {\partial _\mu }{\gamma _{1{{\bf{X}}_ \bot }}}\cdot d{\bf{X}}$. Eventually, the new 1-form of $O(\varepsilon^{1+\sigma})$ is
\begin{equation}\label{a74}
\begin{array}{l}
{\Gamma _{1 + \sigma }} = d{S_{1 + \sigma }} + \left( {{\varepsilon ^{ - \sigma }}{\bf{g}}_1^{\bf{X}}\cdot\nabla {\phi _\sigma } - g_\sigma ^\mu {\partial _\mu }{\gamma _{1t}}} \right)dt  \\
 + \left[ {{\bf{g}}_{1 + \sigma }^{\bf{X}} \times {\bf{B}}\left( {\bf{X}} \right) - \frac{1}{2}g_\sigma ^\mu {\partial _\mu }{\gamma _{1{{\bf{X}}_ \bot }}}} \right]\cdot d{\bf{X}},
\end{array}
\end{equation}
where only $-\mu B(\mathbf{X})$ in $\gamma_{1t}$ contributes. The function of multiplying a factor $\varepsilon^{-\sigma}$ is to make $\varepsilon^{-\sigma} \nabla \phi_\sigma$ be of order $O(1)$, since $\nabla \phi_\sigma$ is of order $\varepsilon^{-\sigma}$. Then $\Gamma_{1+\sigma}$ is of order $O(1)$.

The cancellation equations in Eq.(\ref{a74}) are
\begin{equation}\label{a76}
\varepsilon^{-\sigma}\mathbf{g}_1^{\bf{X}}\cdot\nabla {\phi _\sigma } = g_\sigma ^\mu {\partial _\mu }{\gamma _{1t}},
\end{equation}
\begin{equation}\label{a77}
{\bf{g}}_{1 + \sigma }^{\bf{X}} \times {\bf{B}}\left( {\bf{X}} \right) = \frac{1}{2}g_\sigma ^\mu {\partial _\mu }{\gamma _{1{{\bf{X}}_ \bot }}},
\end{equation}
which lead to the solutions
\begin{equation}\label{a78}
g_\sigma ^\mu  = \frac{{{\varepsilon ^{ - \sigma }}{\bf{g}}_1^{\bf{X}}\cdot\nabla {\phi _\sigma }}}{{B\left( {\bf{X}} \right)}}.
\end{equation}
\begin{equation}\label{a26}
{\bf{g}}_{1 + \sigma }^{\bf{X}} = \frac{{{\bf{b}} \times g_\sigma ^\mu {\partial _\mu }{\gamma _{1{{\bf{X}}_ \bot }}}}}{{2 B\left( {\bf{X}} \right)}}.
\end{equation}

Eventually, it's derived that ${\Gamma _{1 + \sigma }} = d{S_{1 + \sigma}}$. But it should be remembered that there are higher order $\theta$-coupling terms left in $\mathbf{M}_{1+\sigma}^{\bf{X}}$, which are left to higher order cancellation.

\subsection{Cancelling $\theta$-coupling terms of $O(\varepsilon^{1+\eta})$ in new $\Gamma$}\label{sec5.3.10}

The term of order $O(\varepsilon^{1+\eta})$ in $\mathbf{M}_1^{\mathbf{X}}$ is ${\varepsilon ^{*(1 + \eta )}}{\bf{g}}_1^{\bf{X}}\cdot{\partial _t}{\gamma _{0\alpha {\bf{X}}}}dt$. Just as the cancellation at the order $O(\varepsilon^{1+\sigma})$ in Subsec.(\ref{sec5.3.2}), it needs to introduce generators $\mathbf{g}_{1+\eta}^{\mathbf{X}}, g_{\eta}^{\mu}$. Then, $\Gamma_{1+\eta}$ becomes
\begin{equation}\label{a124}
\begin{array}{l}
{\Gamma _{1 + \eta }} = d{S_{1 + \eta }} + \left( \varepsilon^{-\eta} {{\bf{g}}_1^{\bf{X}}\cdot{\partial _t}{\gamma _{0\alpha {\bf{X}}}} - g_\eta ^\mu {\partial _\mu }{\gamma _{1t}}} \right)dt  \nonumber \\
 + \left[ {{\bf{g}}_{1 + \eta }^{\bf{X}} \times {\bf{B}}\left( {\bf{X}} \right) - \frac{1}{2}g_\eta ^\mu {\partial _\mu }{\gamma _{1{{\bf{X}}_ \bot }}}} \right]\cdot d{\bf{X}}.
\end{array}
\end{equation}
The function of the factor $\varepsilon^{-\eta}$ is the same with that of $\varepsilon^{-\sigma}$.

If the oscillating frequency of $\mathbf{A}_{\alpha}(\mathbf{X},t)$ is of gyrofrequency, ${\bf{g}}_1^{\bf{X}}\cdot{\partial _t}{\gamma _{0\alpha {\bf{X}}}}dt$ in $\mathbf{M}_1^{\mathbf{X}}$ could introduce cyclotron resonance between oscillating frequency and cyclotron frequency of charged particle, and change the magnetic moment. Therefore, if cyclotron resonance happens, this term shouldn't be cancelled, since the composite phase of $\theta$ and rotating angle of circular polarized electric field is not a fast variable anymore. The cyclotron resonance will be studied in an independent paper.

The cancellation equations are
\begin{equation}\label{a125}
\varepsilon^{-\eta}{\bf{g}}_1^{\bf{X}}\cdot{\partial _t}{\gamma _{0\alpha {\bf{X}}}} = g_\eta ^\mu {\partial _\mu }{\gamma _{1t}},
\end{equation}
\begin{equation}\label{a126}
{\bf{g}}_{1 + \eta }^{\bf{X}} \times {\bf{B}}\left( {\bf{X}} \right) = \frac{1}{2}g_\eta ^\mu {\partial _\mu }{\gamma _{1{{\bf{X}}_ \bot }}},
\end{equation}
which give solution of ${\bf{g}}_{1 + \eta }^{\bf{X}}, g_\eta ^\mu $
\begin{equation}\label{a160}
g_\eta ^\mu  = \frac{{{\varepsilon ^{ - \eta }}{\bf{g}}_1^{\bf{X}}\cdot{\partial _t}{{\bf{A}}_\alpha }}}{{B\left( {\bf{X}} \right)}},
\end{equation}
\begin{equation}\label{a161}
{\bf{g}}_{1 + \eta }^{\bf{X}} = \frac{{g_\eta ^\mu {\bf{b}} \times {\partial _\mu }{\gamma _{1 \bot }}}}{{2 B\left( {\bf{X}} \right)}}.
\end{equation}
Eventually, ${\Gamma _{1 + \eta }}=dS_{1+\eta}$ is derived.

\subsection{Cancelling $\theta$-coupling terms of $O(\varepsilon^{1+\alpha})$ in new $\Gamma$}\label{sec5.3.3}

The term of $O(\varepsilon^{1+\alpha})$ in $\mathbf{M}_1^{\mathbf{X}}$ is ${\bf{g}}_1^{\bf{X}} \times \mathbf{B}_{\alpha}$,  where $\mathbf{B}_{\alpha}=\left( {\nabla  \times {\Gamma _{0\alpha{\bf{X}}}}} \right)$. As pointed out before, ${\nabla  \times {\Gamma _{0\alpha{\bf{X}}}}}$ is a physical term and of order $\varepsilon^\alpha$. To cancel the perpendicular part of this term,  generator $\mathbf{g}_{1+\alpha}^{\mathbf{X}}$ is introduced. The introduction of $\mathbf{g}_{1+\alpha}^{\mathbf{X}}$ generates a set of linear terms $\mathbf{M}_{1+\alpha}^{\bf{X}}$ similar to $\mathbf{M}_{1}^{\bf{X}}$. The first term in $\mathbf{M}_{1+\alpha}^{\bf{X}}$, which is of lowest order, is used to cancel the perpendicular part.

To cancel the parallel part of this term, the generator $g^U_{\alpha}$ is introduced and the lowest order terms generated are
\begin{equation}\label{b13}
{\bf{M}}_\alpha ^U =  -\varepsilon^{1+\alpha} g_\alpha ^U{\partial _U}{\gamma _{1{\bf{X}}\parallel }}\cdot d{\bf{X}} - \varepsilon^{1+\alpha} g_\alpha ^U{\partial _U}{\gamma _{1t}} dt.
\end{equation}
The first term is used to cancel the parallel part. But the second term needs to be canceled by introducing a generator $g_{\alpha}^{\mu}$, the lowest order terms generated by which is recorded as $\mathbf{M}_{\alpha}^{\mu}$.
\begin{equation}\label{b1}
\begin{array}{*{20}{l}}
{{\bf{M}}_\alpha ^\mu  =  - {\varepsilon ^{1 + \alpha }}{L_{g_\alpha ^\mu }}{\Gamma _1} - \frac{1}{2}{\varepsilon ^{1 + \alpha }}g_\alpha ^\mu {\partial _\mu }{\gamma _{1{{\bf{X}}_ \bot }}}\cdot d{\bf{X}}}\\
{ =  - {\varepsilon ^{1 + \alpha }}g_\alpha ^\mu {\partial _\mu }{\gamma _{1t}}dt - \frac{1}{2}{\varepsilon ^{1 + \alpha }}g_\alpha ^\mu {\partial _\mu }{\gamma _{1{{\bf{X}}_ \bot }}}\cdot d{\bf{X}}}
\end{array}
\end{equation}
The first term of $\mathbf{M}_{\alpha}^{\mu}$ is used to cancel the second term in ${\bf{M}}_\alpha ^U$. The left term can be cancelled by the first term of $\mathbf{M}_{1+\alpha}^{\bf{X}}$.
Then, the new $\Gamma$ of $O(\varepsilon^{1+\alpha})$ is
\begin{equation}\label{a75}
\begin{array}{l}
{\Gamma _{1 + \alpha }} = d{S_{1 + \alpha }} + \left[ \varepsilon^{-\alpha} {{\bf{g}}_1^{\bf{X}} \times {{\bf{B}}_\alpha } + {\bf{g}}_{1 + \alpha }^{\bf{X}} \times {\bf{B}}\left( {\bf{X}} \right) - g_\alpha ^U{\partial _U}{\gamma _{1{\bf{X}}\parallel }} - \frac{1}{2}g_\alpha ^\mu {\partial _\mu }{\gamma _{1{{\bf{X}}_ \bot }}}} \right]\cdot d{\bf{X}}\\
 - \left[ {g_\alpha ^U{\partial _U}{\gamma _{1t}} + g_\alpha ^\mu {\partial _\mu }{\gamma _{1t}}} \right]dt.
\end{array}
\end{equation}
The function of the factor $\varepsilon^{-\alpha}$ is the same with that of $\varepsilon^{-\sigma}$.

The cancellation equations in Eq.(\ref{a75}) are
\begin{equation}\label{a79}
\varepsilon^{-\alpha} {\left( {{\bf{g}}_1^{\bf{X}} \times {{\bf{B}}_\alpha }} \right)_ \bot } - \frac{1}{2}g_\alpha ^\mu {\partial _\mu }{\gamma _{1{{\bf{X}}_ \bot }}} =  - {\bf{g}}_{1 + \alpha }^{\bf{X}} \times {\bf{B}}\left( {\bf{X}} \right),
\end{equation}
\begin{equation}\label{a80}
\varepsilon^{-\alpha}{\left( {{\bf{g}}_1^{\bf{X}} \times {{\bf{B}}_\alpha }} \right)_\parallel } = g_\alpha ^U{\partial _U}{\gamma _{1{\bf{X}}\parallel }},
\end{equation}
\begin{equation}\label{a81}
g_\alpha ^U{\partial _U}{\gamma _{1t}} =  - g_\alpha ^\mu {\partial _\mu }{\gamma _{1t}}.
\end{equation}
The solutions are
\begin{equation}\label{a82}
g_\alpha ^U = \frac{\varepsilon^{-\alpha}{{{\left( {{\bf{g}}_1^{\bf{X}} \times {{\bf{B}}_\alpha }} \right)}_\parallel }}}{{{\partial _U}{\gamma _{1{\bf{X}}\parallel }}}} =\varepsilon^{-\alpha} \sqrt {\frac{{2\mu }}{{B\left( {\bf{X}} \right)}}} {\widehat {\bf{v}}_ \bot }\cdot{{\bf{B}}_\alpha }
\end{equation}
where ${\bf{b}} \times \hat{\rho}  =  - {\hat {\bf{v}}_ \bot }$ is applied,
\begin{equation}\label{a83}
g_\alpha ^\mu  =  - \frac{\varepsilon^{-\alpha}{g_\alpha ^U{\partial _U}{\gamma _{1t}}}}{{{\partial _\mu }{\gamma _{1t}}}} = \frac{\varepsilon^{-\alpha} U}{{B({\bf{X}})}}\sqrt {\frac{{2\mu }}{{B\left( {\bf{X}} \right)}}} {\widehat {\bf{v}}_ \bot }\cdot{{\bf{B}}_\alpha },
\end{equation}
\begin{equation}\label{a84}
{\bf{g}}_{1 + \alpha }^{\bf{X}} = \frac{\varepsilon^{-\alpha}{\left[ {\left( {{\bf{b}}\cdot{{\bf{B}}_\alpha }} \right){\bf{g}}_1^{\bf{X}} - \frac{U}{{2 B({\bf{X}})}}\left( {{{\widehat {\bf{v}}}_ \bot }\cdot{{\bf{B}}_\alpha }} \right)\hat \rho } \right]}}{{B\left( {\bf{X}} \right)}}.
\end{equation}

Eventually, it's derived that ${\Gamma _{1 + \alpha }} = d{S_{1 + \alpha }}$.  But it should be remembered that there are higher order $\theta$-coupling terms left in $\mathbf{M}_{1+\alpha}^{\bf{X}}$, which are left to higher order cancellation.

\subsection{Canceling $\theta$-coupling terms of $O(\varepsilon^2)$ in new $\Gamma$}\label{sec5.3.5}

The second order is written as
\begin{equation}\label{af53}
\begin{array}{l}
{\Gamma _2} = d{S} + \frac{1}{2}{ {{L^2_{{\bf{g}}_1^{\bf{X}}}}} }{\gamma _0} - {L_{{\bf{g}}_1^{\bf{X}}}}{\gamma _1} + {(?)_2}\\
 = d{S} - \frac{1}{2}L_{{\bf{g}}_1^{\bf{X}}}^2{\gamma _0} - {L_{{\bf{g}}_1^{\bf{X}}}}{\Gamma _1}+ {(?)_2}.
\end{array}
\end{equation}
Here, symbol $(?)_2$ represents terms generated by newly introduced generators in $\Gamma_2$.  The already known two terms in Eq.(\ref{af53}) are
%
\begin{eqnarray}\label{af45}
{L_{{\bf{g}}_1^{\bf{X}}}}{\Gamma _1} =  - U{\bf{g}}_1^{\bf{X}} \times \left( {\nabla  \times {\bf{b}}} \right)\cdot d{\bf{X}} + {\bf{g}}_1^{\bf{X}}\cdot\nabla {\Gamma _{1t}}dt+dS
\end{eqnarray}
\begin{equation}\label{af54}
\begin{array}{l}
\frac{1}{2}{\left( {{L_{g_1^{\bf{X}}}}} \right)^2}{\gamma _0} = \frac{1}{2}{\bf{g}}_1^{\bf{X}} \times \left( {\nabla  \times \left( {{\bf{g}}_1^{\bf{X}} \times {{\bf{B}}_0}} \right)} \right)\cdot d{\bf{X}}\\
  +\frac{1}{2}{\bf{g}}_1^{\bf{X}}\cdot{\partial _\mu }\left( {{\bf{g}}_1^{\bf{X}} \times {{\bf{B}}_0}} \right)d\mu  + \frac{1}{2}{\bf{g}}_1^{\bf{X}}\cdot{\partial _\theta }\left( {{\bf{g}}_1^{\bf{X}} \times {{\bf{B}}_0}} \right)d\theta
\end{array}.
\end{equation}

All known $\mathbf{X}$ component in Eq.(\ref{af53}) is recorded as $\Gamma_{o2\bf{X}}$, which is the summation of the first terms of RHS of Eq.(\ref{af45}) and  Eq.(\ref{af54}). All known $t$ component in Eq.(\ref{af53}) is recorded as $\Gamma_{o2t}$. To cancel perpendicular part $[\Gamma_{o2\bf{X}}]_\perp$, it needs to introduce a generator field ${\bf{g}}_2^{\;{\bf{X}}}$, which generates a linear term set ${\bf{M}}_2^{\bf{X}}$  similar to ${\bf{M}}_1^{\bf{X}}$ given in Eq.(\ref{a18}). To cancel parallel part $[\Gamma_{o2\bf{X}}]_\parallel$, it needs to introduce a generator field $g_1^U$, which generates a linear term set ${\bf{M}}_1^U $. To cancel the $t$ component, it needs to introduce a generator $g_1^\mu$, which generates a linear terms set ${\bf{M}}_1^\mu$.

Then, it's derived that
\begin{equation}\label{af58}
{(?)_2} =  - \left( {{L_{{\bf{g}}_2^{\bf{X}}}}{\gamma _0}} \right)\left( {\bf{Z}} \right) + \frac{1}{2}\left[ {{L_{g_1^\mu }}{L_{{\bf{g}}_1^{\bf{X}}}}{\gamma _0}} \right]\left( {\bf{Z}} \right) - \left( {{L_{g_1^U}} + {L_{g_1^\mu }}} \right){\gamma _1}\left( {\bf{Z}} \right).
\end{equation}
Now substituting Eqs.(\ref{af45},\ref{af54},\ref{af58}) back into Eq.(\ref{af53}), the cancellation equations for $\bf{X}$ and $t$ components are
\begin{equation}\label{af55}
{\left[ {{\Gamma _{o2{\bf{X}}}} - \frac{1}{2}\left( {{L_{g_1^\mu }}{L_{{\bf{g}}_1^{\bf{X}}}}{\gamma _0}} \right)} \right]_ \bot } = {\bf{g}}_2^{\bf{X}} \times {\bf{B}},
\end{equation}
\begin{equation}\label{af56}
{\left[ {{\Gamma _{o2{\bf{X}}}}} \right]_\parallel } = g_1^U{\bf{b}},
\end{equation}
\begin{equation}\label{af57}
g_1^U{\partial _U}{\gamma _{1t}} - {\bf{g}}_1^{\bf{X}}\cdot\nabla {\gamma _{1t}} =  - g_1^\mu {\partial _\mu }{\gamma _{1t}}.
\end{equation}

After the cancellation, $\Gamma_2$ becomes
\begin{eqnarray}\label{af92}
{\Gamma _2} =&& dS  - \frac{1}{2}{\bf{g}}_1^{\bf{X}}\cdot{\partial _\theta }\left( {{\bf{g}}_1^{\bf{X}} \times {\bf{B}}} \right)d\theta  \nonumber \\
&&- \frac{1}{2}{\bf{g}}_1^{\bf{X}}\cdot{\partial _\mu }\left( {{\bf{g}}_1^{\bf{X}} \times {\bf{B}}} \right)d\mu
\end{eqnarray}
It's noticed that ${\bf{g}}_1^{\bf{X}}\cdot{\partial _\mu }\left( {{\bf{g}}_1^{\bf{X}} \times {\bf{B}}} \right) = 0$ and $-\frac{1}{2}{\bf{g}}_1^{\bf{X}}\cdot{\partial _\theta }\left( {{\bf{g}}_1^{\bf{X}} \times {{\bf{B}}}} \right) =  \mu $, then, $\Gamma_2$ is simplified to be
\begin{equation}\label{af93}
\Gamma_2=dS+\mu d\theta
\end{equation}

\subsection{Effect of spatial gradient and frequency of $\mathbf{B}_{\alpha}(\mathbf{X},t)$}\label{sec6.6}

Now, let's consider the effect of spatial gradient and frequency of perturbation. It's first to consider the effect of $\mathbf{B}_{\alpha}(\mathbf{X},t)$. It's noted from Subsecs.(\ref{sec5.3.1}-\ref{sec5.3.3}) that linear terms doesn't contribute to the effect of spatial gradient and frequency of $\mathbf{B}_{\alpha}(\mathbf{X},t)$, as they are all cancelled. This effect may come from nonlinear term.

Based on $\Gamma_{1+\alpha}$ given in Eq.(\ref{a75}), the lowest-order nonlinear terms which may include the effect of spatial gradient and frequency of $\mathbf{B}_{\alpha}(\mathbf{X},t)$ are recorded as $\Gamma_{2+\alpha-\beta}$ and $\Gamma_{2+\alpha-\tau}$, respectively. By expanding the 1-form formula $\left[ {\exp ( - \varepsilon {L_{{\bf{g}}_1^{\bf{X}}}} - {\varepsilon ^\alpha }{L_{g_\alpha ^\mu }} - {\varepsilon ^\alpha }{L_{g_\alpha ^U}} -  \cdots )\gamma } \right]\left( {\bf{Z}} \right)$, it's derived  that
\begin{equation}\label{a128}
\begin{array}{l}
\varepsilon^{2+\alpha-\beta}{\Gamma _{2 + \alpha  - \beta }} = d{S_{2 + \alpha  - \beta }} + \frac{1}{2}\varepsilon^{2+\alpha} {\left( {L_{{\bf{g}}_1^{\bf{X}}}}{\Gamma _{1 + \alpha }}\right)_{2 + \alpha  - \beta }} \\
- \frac{1}{6}\varepsilon^{2+\alpha}{\left( {{L_{{\bf{g}}_1^{\bf{X}}}}\left( {{L_{g_\alpha ^\mu }} + {L_{g_\alpha ^U}}} \right){L_{{\bf{g}}_1^{\bf{X}}}}{\gamma _0}} \right)_{2 + \alpha  - \beta }},
\end{array}
\end{equation}
\begin{equation}\label{a129}
\begin{array}{l}
\varepsilon^{2+\alpha-\beta}{\Gamma _{2 + \alpha  - \tau }} = d{S_{2 + \alpha  - \tau }} + \frac{1}{2}\varepsilon^{2+\alpha}{\left( {L_{{\bf{g}}_1^{\bf{X}}}}{\Gamma _{1 + \alpha }} \right)_{2 + \alpha  - \tau }} \\
- \frac{1}{6}\varepsilon^{2+\alpha}{\left( {{L_{{\bf{g}}_1^{\bf{X}}}}\left( {{L_{g_\alpha ^\mu }} + {L_{g_\alpha ^U}}} \right){L_{{\bf{g}}_1^{\bf{X}}}}{\gamma _0}} \right)_{2 + \alpha  - \tau }},
\end{array}
\end{equation}
where subscripts $2+\alpha-\beta, 2+\alpha-\tau$ indicate that the parts of order $O(\varepsilon^{2+\alpha-\beta})$ and $O(\varepsilon^{2+\alpha-\tau})$ included by the terms, respectively. $g_{\alpha}^U,g_{\alpha}^{\mu}$ are already given in Eqs.(\ref{a82},\ref{a83}) and they include cross product between $\mathbf{B}_{\alpha}$ and $\mathbf{g}_1^{\bf{X}}$. In Eq.(\ref{a128}), factor $-\beta$ comes from the action of gradient operator over perturbation, i.e. the gradient operator in $\mathbf{g}_1^{\mathbf{X}}\times \nabla_{\beta} \times$ generated by $L_{\mathbf{g}_1^{\mathbf{X}}}$ act over $\mathbf{B}_\alpha$. Here, subscript $\beta$ means the operator $\nabla$ operating directly on the perturbation. And factor $-\tau$ comes from the action of operator $-\mathbf{g}_1^{\mathbf{X}}\cdot \partial_{t,\tau}$ included in $L_{\mathbf{g}_1^{\mathbf{X}}}$ over $\mathbf{B}_\alpha$.

In Eq.(\ref{a128}), quadratic terms of order $O(\varepsilon^{2+\alpha-\beta})$ are completely included by ${L_{{\bf{g}}_1^{\bf{X}}}}{\Gamma _{1 + \alpha }}$. The only cubic term is the third term on RHS of Eq.(\ref{a128}). No linear term exists of order $\Gamma_{2+\alpha-\beta}$. The same constitution is applied to Eq.(\ref{a129}).

The second term in Eqs.(\ref{a128},\ref{a129}) is a gauge term or full differential term. The third term  in both equations includes spatial gradient and frequency effect, respectively, but it is a product of three generators that it only include $\theta$-coupling term, which should be killed. The third terms of RHS of Eq.(\ref{a128}) and (\ref{a129}) are in $\mathbf{X}$ and $t$ component, respectively. They can be cancelled based on the linear cancellation rule.

Therefore, it's necessary to go to next order to find terms including effect of spatial gradient and frequency, and three kinds of 1-form are derived
\begin{equation}\label{a130}
\begin{array}{l}
\varepsilon^{3 + \alpha  - 2\beta}{\Gamma _{3 + \alpha  - 2\beta }} = d{S_{3 + \alpha  - 2\beta }} - \frac{1}{6}\varepsilon^{3+\alpha} {\left( {{L_{{\bf{g}}_1^{\bf{X}}}}{\Gamma _{2 + \alpha-\beta }}} \right)_{3 + \alpha  - 2\beta }}\\
 + \frac{1}{{24}}\varepsilon^{3+\alpha} {\left( {{L_{{\bf{g}}_1^{\bf{X}}}}{L_{{\bf{g}}_1^{\bf{X}}}}\left( {{L_{g_\alpha ^\mu }} + {L_{g_\alpha ^U}}} \right){L_{{\bf{g}}_1^{\bf{X}}}}{\gamma _0}} \right)_{3 + \alpha  - 2\beta }}
\end{array}
\end{equation}
\begin{equation}\label{a131}
\begin{array}{l}
\varepsilon^{3 + \alpha  - \beta-\tau}{\Gamma _{3 + \alpha  - \beta  - \tau }} = d{S_{3 + \alpha  - \beta  - \tau }} - \frac{1}{6}\varepsilon^{3+\alpha} {\left( {{L_{{\bf{g}}_1^{\bf{X}}}}{\Gamma _{2 + \alpha-\beta }}} \right)_{3 + \alpha  - \beta  - \tau }}\\
 + \frac{1}{{24}}\varepsilon^{3+\alpha} {\left( {{L_{{\bf{g}}_1^{\bf{X}}}}{L_{{\bf{g}}_1^{\bf{X}}}}\left( {{L_{g_\alpha ^\mu }} + {L_{g_\alpha ^U}}} \right){L_{{\bf{g}}_1^{\bf{X}}}}{\gamma _0}} \right)_{3 + \alpha  - \beta  - \tau }}
\end{array}
\end{equation}
\begin{equation}\label{a132}
\begin{array}{l}
\varepsilon^{3 + \alpha  - 2\tau}{\Gamma _{3 + \alpha  - 2\tau }} = d{S_{3 + \alpha  - 2\tau }} - \frac{1}{6}\varepsilon^{3+\alpha}{\left( {{L_{{\bf{g}}_1^{\bf{X}}}}{\Gamma _{2 + \alpha-\beta }}} \right)_{3 + \alpha  - 2\tau }}\\
 + \frac{1}{{24}}\varepsilon^{3+\alpha} {\left( {{L_{{\bf{g}}_1^{\bf{X}}}}{L_{{\bf{g}}_1^{\bf{X}}}}\left( {{L_{g_\alpha ^\mu }} + {L_{g_\alpha ^U}}} \right){L_{{\bf{g}}_1^{\bf{X}}}}{\gamma _0}} \right)_{3 + \alpha  - 2\tau }} \\
 =d{S_{3 + \alpha  - 2\tau }}
\end{array}
\end{equation}
There terms only include cubic and quartic terms. For such order, no quadratic and linear terms exists.

On RHS of Eqs.(\ref{a130}-\ref{a132}), the first two terms for each question is a full differential term. For the third term, the action of $L_{g_{\alpha}^{U}}$ over $L_{\mathbf{g}_1^{\mathbf{X}}}\gamma_0$ is zero, then, only $L_{g_{\alpha}^{\mu}}$ exists. The third term in Eq.(\ref{a130}) is in $\mathbf{X}$ components. Its $\theta$-coupling part can be cancelled according to the linear cancellation rule. The non-$\theta$-coupling part is left as term of effect of spatial gradient of $\mathbf{B}_\alpha$ and is recorded as $\left\langle {{\Gamma _{3 + \alpha  - 2\beta }}} \right\rangle_{\mathbf{X}} $. The third term in Eq.(\ref{a131}) includes a term in $\mathbf{X}$ component and a term in $t$ component, the non-$\theta$-coupling parts of which are recorded as ${\left\langle {{\Gamma _{3 + \alpha  - \beta  - \tau }}} \right\rangle _{\bf{X}}}$ and ${\left\langle {{\Gamma _{3 + \alpha  - \beta  - \tau }}} \right\rangle _t}$, respectively. The third term in Eq.(\ref{a132}) equals zero, due to the following reason. The operator ${L_{{\bf{g}}_1^{\bf{x}}}}(\cdots)$ in this term operates like ${\bf{g}}_1^{\bf{x}}\cdot{\partial _{t,\tau}}(\cdots)dt$, where $\partial_{t,\tau}$ denotes $\partial_t$ directly acting upon perturbation potential. The twice repeating like ${\bf{g}}_1^{\bf{x}}\cdot{\partial _{t,\tau}}\left( {{\bf{g}}_1^{\bf{x}}\cdot{\partial _{t,\tau}}\left(  \cdots  \right)dt} \right)$ equals zero according to Eq.(\ref{vp7}). All these quantities will be calculated in \ref{app2}

\subsection{The effect of spatial gradient and frequency of $\nabla \phi_\sigma $}\label{sec6.7}

Based on $\Gamma_{1+\sigma}$ given in Eq.(\ref{a75}), the lowest-order nonlinear terms which may include the effect of spatial gradient and frequency of $\nabla \phi_\sigma$ are recorded as $\Gamma_{2+\sigma-\beta}$ and $\Gamma_{2+\sigma-\tau}$, respectively. By expanding the 1-form formula $\left[ {\exp ( - \varepsilon {L_{{\bf{g}}_1^{\bf{X}}}} - {\varepsilon ^\sigma }{L_{g_\sigma ^\mu }} -  \cdots )\gamma } \right]\left( {\bf{Z}} \right)$, it's derived  that
\begin{equation}\label{a134}
\begin{array}{l}
\varepsilon^{2 + \sigma  - \beta}{\Gamma _{2 + \sigma  - \beta }} = d{S_{2 + \sigma  - \beta }} +\frac{1}{2}\varepsilon^{2+\sigma} {\left( {L_{{\bf{g}}_1^{\bf{X}}}}{\Gamma _{1 + \sigma }} \right)_{2 + \sigma  - \beta }}  \\
- \frac{1}{6}\varepsilon^{2+\sigma} {\left( {{L_{{\bf{g}}_1^{\bf{X}}}}{L_{g_\sigma ^\mu }}{L_{{\bf{g}}_1^{\bf{X}}}}{\gamma _0}} \right)_{2 + \sigma  - \beta }},
\end{array}
\end{equation}
\begin{equation}\label{a135}
\begin{array}{l}
\varepsilon^{2 + \sigma  - \tau}{\Gamma _{2 + \sigma  - \tau }} = d{S_{2 + \sigma  - \tau }} + \frac{1}{2}\varepsilon^{2+\sigma}{\left({L_{{\bf{g}}_1^{\bf{X}}}}{\Gamma _{1 + \sigma }} \right)_{2 + \sigma  - \tau }}  \\
- \frac{1}{6}\varepsilon^{2+\sigma} {\left( {{L_{{\bf{g}}_1^{\bf{X}}}}{L_{g_\sigma ^\mu }}{L_{{\bf{g}}_1^{\bf{X}}}}{\gamma _0}} \right)_{2 + \sigma  - \tau }}.
\end{array}
\end{equation}
The meaning of subscripts $2 + \sigma  - \beta$ and $2 + \sigma  - \tau $ are the same with $2 + \alpha  - \beta$ and $2 + \alpha  - \tau$ given in Subsec.(\ref{sec6.6}).
The third term on RHS of Eqs.(\ref{a134}) and (\ref{a135}) only includes $\theta$-coupling term and should be killed. Just as the discussion for Eqs.(\ref{a128}) and (\ref{a129}), Eqs.(\ref{a134}) and (\ref{a135}) doesn't contribute term of the effect of spatial gradient and frequency of $\nabla \phi_{\sigma}$.
We need go to next order terms as follows
\begin{equation}\label{a136}
\begin{array}{l}
\varepsilon^{3 + \sigma  - 2\beta} {\Gamma _{3 + \sigma  - 2\beta }} = d{S_{3 + \sigma  - 2\beta }} - \frac{1}{6}\varepsilon^{3+\sigma} {\left( {{L_{{\bf{g}}_1^{\bf{X}}}}{\Gamma _{2 + \sigma  - \beta }}} \right)_{3 + \sigma  - 2\beta }}\\
 + \frac{1}{{24}}\varepsilon^{3+\sigma} {\left( {{L_{{\bf{g}}_1^{\bf{X}}}}{L_{{\bf{g}}_1^{\bf{X}}}}{L_{g_\sigma ^\mu }}{L_{{\bf{g}}_1^{\bf{X}}}}{\gamma _0}} \right)_{3 + \sigma  - 2\beta }},
\end{array}
\end{equation}
\begin{equation}\label{a137}
\begin{array}{l}
\varepsilon^{3 + \sigma  - \beta  - \tau}{\Gamma _{3 + \sigma  - \beta  - \tau }} = d{S_{3 + \sigma  - \beta  - \tau }} - \frac{1}{6}\varepsilon^{3+\sigma} {\left( {{L_{{\bf{g}}_1^{\bf{X}}}}{\Gamma _{2 + \sigma  - \beta }}} \right)_{3 + \sigma  - \beta  - \tau }}\\
 + \frac{1}{{24}}\varepsilon^{3+\sigma} {\left( {{L_{{\bf{g}}_1^{\bf{X}}}}{L_{{\bf{g}}_1^{\bf{X}}}}{L_{g_\sigma ^\mu }}{L_{{\bf{g}}_1^{\bf{X}}}}{\gamma _0}} \right)_{3 + \sigma  - \beta  - \tau }},
\end{array}
\end{equation}
\begin{equation}\label{a138}
\begin{array}{l}
\varepsilon^{3 + \sigma  - 2\tau} {\Gamma _{3 + \sigma  - 2\tau }} = d{S_{3 + \sigma  - 2\tau }} - \frac{1}{6}\varepsilon^{3+\sigma}{\left( {{L_{{\bf{g}}_1^{\bf{X}}}}{\Gamma _{2 + \sigma  - \beta }}} \right)_{3 + \sigma  - 2\tau }}\\
 + \frac{1}{{24}}\varepsilon^{3+\sigma} {\left( {{L_{{\bf{g}}_1^{\bf{X}}}}{L_{{\bf{g}}_1^{\bf{X}}}}{L_{g_\sigma ^\mu }}{L_{{\bf{g}}_1^{\bf{X}}}}{\gamma _0}} \right)_{3 + \sigma  - 2\tau }}=d{S_{3 + \sigma  - 2\tau }}.
\end{array}
\end{equation}

On RHS of Eqs.(\ref{a136}-\ref{a138}), the first two terms for each equation is a full differential term. The third term in Eq.(\ref{a136}) is in $\mathbf{X}$ components. It's $\theta$-coupling part can be cancelled according to the linear cancellation rule. The non-$\theta$-coupling part is left as term of effect of spatial gradient of $\nabla \phi_{\sigma}$ and is recorded as $\left\langle {{\Gamma _{3 + \sigma  - 2\beta }}} \right\rangle_{\mathbf{X}} $. The third term in Eq.(\ref{a137}) includes a term in $\mathbf{X}$ component and a term in $t$ component, the non-$\theta$-coupling parts of which are recorded as ${\left\langle {{\Gamma _{3 + \sigma  - \beta  - \tau }}} \right\rangle _{\bf{X}}}$ and ${\left\langle {{\Gamma _{3 + \sigma  - \beta  - \tau }}} \right\rangle _t}$, respectively. The third term in Eq.(\ref{a138}) equals zero. All these quantities will be calculated in \ref{app2}.

\subsection{The effect of spatial gradient and frequency of ${\partial _t} \mathbf{A}_{\alpha}$}\label{sec6.8}

Based on $\Gamma_{1+\sigma}$ given in Eq.(\ref{a75}), the lowest-order nonlinear terms which may include the effect of spatial gradient and frequency of ${\partial _t} \mathbf{A}_{\alpha}$ are recorded as $\Gamma_{2+\eta-\beta}$ and $\Gamma_{2+\eta-\tau}$, respectively. By expanding the 1-form formula $\left[ {\exp ( - \varepsilon {L_{{\bf{g}}_1^{\bf{X}}}} - {\varepsilon ^{\eta} }{L_{g_\eta ^\mu }} -  \cdots )\gamma } \right]\left( {\bf{Z}} \right)$, it's derived  that
\begin{eqnarray}\label{a139}
\varepsilon^{2 + \eta  - \beta} {\Gamma _{2 + \eta  - \beta }} = d{S_{2 + \eta  - \beta }} + \varepsilon^{2+\eta} {\left( {{L_{{\bf{g}}_1^{\bf{X}}}}{\Gamma _{1 + \eta }}} \right)_{2 + \eta  - \beta }} \nonumber  \\
- \frac{1}{6}\varepsilon^{2+\eta} {\left( {{L_{{\bf{g}}_1^{\bf{X}}}}{L_{g_\eta ^\mu }}{L_{{\bf{g}}_1^{\bf{X}}}}{\gamma _0}} \right)_{2 + \eta  - \beta }},
\end{eqnarray}
\begin{eqnarray}\label{a140}
\varepsilon^{2 + \eta  - \tau} {\Gamma _{2 + \eta  - \tau }} = d{S_{2 + \eta  - \tau }} + \varepsilon^{2+\eta}{\left( {{L_{{\bf{g}}_1^{\bf{X}}}}{\Gamma _{1 + \eta }}} \right)_{2 + \eta  - \tau }} \nonumber  \\
- \frac{1}{6}\varepsilon^{2+\eta} {\left( {{L_{{\bf{g}}_1^{\bf{X}}}}{L_{g_\eta ^\mu }}{L_{{\bf{g}}_1^{\bf{X}}}}{\gamma _0}} \right)_{2 + \eta  - \tau }}.
\end{eqnarray}
Just as the explanation given in Subsecs.(\ref{sec6.6}) and (\ref{sec6.7}), Eqs.(\ref{a139}) and (\ref{a140}) don't contribute to the effect of spatial gradient and frequency of perturbation. We need to explore the following next order terms
\begin{equation}\label{a141}
\begin{array}{l}
\varepsilon^{3 + \eta  - 2\beta} {\Gamma _{3 + \eta  - 2\beta }} = d{S_{3 + \eta  - 2\beta }} - \frac{1}{6}\varepsilon^{3+\eta} {\left( {{L_{{\bf{g}}_1^{\bf{X}}}}{\Gamma _{2 + \eta  - \beta }}} \right)_{3 + \eta  - 2\beta }}\\
 + \frac{1}{{24}}\varepsilon^{3+\eta} {\left( {{L_{{\bf{g}}_1^{\bf{X}}}}{L_{{\bf{g}}_1^{\bf{X}}}}{L_{g_\eta ^\mu }}{L_{{\bf{g}}_1^{\bf{X}}}}{\gamma _0}} \right)_{3 + \eta  - 2\beta }},
\end{array}
\end{equation}
\begin{equation}\label{a142}
\begin{array}{l}
\varepsilon^{3 + \eta  - \beta-\tau}{\Gamma _{3 + \eta  - \beta  - \tau }} = d{S_{3 + \eta  - \beta  - \tau }} - \frac{1}{6}\varepsilon^{3+\eta} {\left( {{L_{{\bf{g}}_1^{\bf{X}}}}{\Gamma _{2 + \eta  - \beta }}} \right)_{3 + \eta  - \beta  - \tau }}\\
 + \frac{1}{{24}}\varepsilon^{3+\eta} {\left( {{L_{{\bf{g}}_1^{\bf{X}}}}{L_{{\bf{g}}_1^{\bf{X}}}}{L_{g_\eta ^\mu }}{L_{{\bf{g}}_1^{\bf{X}}}}{\gamma _0}} \right)_{3 + \eta  - \beta  - \tau }},
\end{array}
\end{equation}
\begin{equation}\label{a143}
\begin{array}{l}
\varepsilon^{3 + \eta -2\tau}{\Gamma _{3 + \eta  - 2\tau }} = d{S_{3 + \eta  - 2\tau }} - \frac{1}{6}\varepsilon^{3+\eta} {\left( {{L_{{\bf{g}}_1^{\bf{X}}}}{\Gamma _{2 + \eta  - \beta }}} \right)_{3 + \eta  - 2\tau }}\\
 + \frac{1}{{24}}\varepsilon^{3+\eta} {\left( {{L_{{\bf{g}}_1^{\bf{X}}}}{L_{{\bf{g}}_1^{\bf{X}}}}{L_{g_\eta ^\mu }}{L_{{\bf{g}}_1^{\bf{X}}}}{\gamma _0}} \right)_{3 + \eta  - 2\tau }}=d{S_{3 + \eta  - 2\tau }}.
\end{array}
\end{equation}

On RHS of Eqs.(\ref{a141}-\ref{a143}), the first two terms for each question is a full differential term. The third term in Eq.(\ref{a141}) is in $\mathbf{X}$ components. It's $\theta$-coupling part can be cancelled according to the linear cancellation rule. The non-$\theta$-coupling part is left as term of effect of spatial gradient of ${\partial _t} \mathbf{A}_{\alpha}$ and is recorded as $\left\langle {{\Gamma _{3 + \eta  - 2\beta }}} \right\rangle_{\mathbf{X}} $. The third term in Eq.(\ref{a142}) includes a term in $\mathbf{X}$ component and a term in $t$ component, the non-$\theta$-coupling parts of which are recorded as ${\left\langle {{\Gamma _{3 + \eta - \beta  - \tau }}} \right\rangle _{\bf{X}}}$ and ${\left\langle {{\Gamma _{3 + \eta - \beta  - \tau }}} \right\rangle _t}$, respectively. The third term in Eq.(\ref{a143}) equals zero. All these quantities will be calculated in \ref{app2}.

\subsection{The restriction condition imposed on perturbation for decoupling  $\theta$ from the remaining degrees of freedom}\label{sec6.9}

The linear cancellation rule shows that to cancel $\theta$-coupling term in $\mathbf{X}$ and $t$ components, we need to introduce new generators which generate new terms. After cancellation, $\theta$-coupling terms of higher order are left. Based on this procedure, a set of generators as $\{{\bf{g}}_1^{\bf{X}},g_1^\mu ,g_1^U,{\bf{g}}_{1 + \sigma }^{\bf{X}},g_\sigma ^\mu ,{\bf{g}}_{1 + \eta }^{\bf{X}},g_\eta ^\mu ,{\bf{g}}_{1 + \alpha }^{\bf{X}},g_\alpha ^\mu ,g_\alpha ^U,{\bf{g}}_2^{\bf{X}}, \cdots \}$ are introduced to the exponential operator $\exp(-\mathbf{E}\cdot L_{\mathbf{g}})$.

The standing condition for linear cancellation rule is that after the cancellation, the left $\theta$-coupling terms are of order higher than that of the original cancelled terms. If the new generated $\theta$-coupling terms are even of order lower than that of original terms, whether linear cancellation rule still stands is suspectable. The factors contributing to lower the order of new generated terms is the spatial gradient and oscillating frequency of perturbation, both of which only involve with generators $\mathbf{g}^{\mathbf{X}}_{\chi}$, as revealed by the two identities ${L_{{\bf{g}}_1^{\bf{X}}}}\left( {{{\bf{A}}_\alpha }\left( {{\bf{X}},t} \right)\cdot d{\bf{X}}} \right) =  - {\bf{g}}_1^{\bf{X}} \times \left( {\nabla  \times {{\bf{A}}_\alpha }\left( {{\bf{X}},t} \right)} \right)\cdot d{\bf{X}} - {\bf{g}}_1^{\bf{X}}\cdot{\partial _t}{\bf{A}}\left( {{\bf{X}},t} \right)dt + dS$ and ${L_{{\bf{g}}_1^{\bf{X}}}}\left( {{\phi _\sigma }\left( {{\bf{X}},t} \right)dt} \right) = {\bf{g}}_1^{\bf{X}}\cdot\nabla {\phi _\sigma }\left( {{\bf{X}},t} \right)dt + dS$. The operator $\nabla$ over perturbation could generate an factor $\varepsilon^{-\beta}$, and $\partial_t$ over perturbation could generate a factor $\varepsilon^{-\tau}$. The lowest order generators among $\mathbf{g}_{\chi}^{\mathbf{X}}$ is $\mathbf{g}_{1}^{\mathbf{X}}$. Therefore, concerned with $\mathbf{g}_{1}^{\mathbf{X}}$, if linear cancellation rule can stands for all spatial gradient and oscillating frequency of perturbation, then linear cancellation rule can stands for all generators.

The new 1-form $\Gamma$ includes a part like
\begin{equation}\label{a144}
\begin{array}{l}
\exp ( - \varepsilon {L_{{\bf{g}}_1^{\bf{X}}}} +  \cdots )({{\bf{A}}_\alpha }({\bf{X}},t)\cdot d{\bf{X}} + \phi ({\bf{X}},t)dt) \\
 = \sum\limits_n {\frac{1}{{n!}}{{\left( { - \varepsilon {L_{{\bf{g}}_1^{\bf{X}}}}} \right)}^n}\left( {{{\bf{A}}_\alpha }({\bf{X}},t)\cdot d{\bf{X}} + \phi ({\bf{X}},t)dt} \right)}  +  \cdots.
\end{array}
\end{equation}
We first consider the effect of spatial gradient and oscillation of $\mathbf{A}_{\alpha}(\mathbf{Z},t)$.
Among all terms including $n$-times product of ${L_{g_\chi ^{}}}$ with $g_{\chi}\in \{ {\bf{g}}_1^{\bf{X}},g_1^\mu ,g_1^U,{\bf{g}}_{1 + \sigma }^{\bf{X}},g_\sigma ^\mu, \cdots \}$, the term ${\left( { - \varepsilon {L_{{\bf{g}}_1^{\bf{X}}}}} \right)^n}\left( {{{\bf{A}}_\alpha }({\bf{X}},t)\cdot d{\bf{X}}} \right)$ can generate lowest order terms. The action of $L_{{\bf{g}}_1^{\bf{X}}}$ on 1-form like $F\left( {{\bf{X}},t} \right)dt$ only generates time component like ${\bf{g}}_1^{\bf{X}}\cdot\nabla F\left( {{\bf{X}},t} \right)dt$. Once a time component $F\left( {{\bf{X}},t} \right)dt$ is generated, the rest action of $L_{{\bf{g}}_1^{\bf{X}}}$ on this time component can only take the form ${\bf{g}}_1^{\bf{X}}\cdot \nabla F\left( {{\bf{X}},t} \right)dt$. The time component can only be generated by the action of $L_{{\bf{g}}_1^{\bf{X}}}$ on 1-form like ${\bf{F}}\left( {{\bf{X}},t} \right)\cdot d{\bf{X}}$. This kind of action generates two component $ - {\bf{g}}_1^{\bf{X}} \times \left( {\nabla  \times {\bf{F}}\left( {{\bf{X}},t} \right)} \right)\cdot d{\bf{X}} - {\bf{g}}_1^{\bf{X}}\cdot{\partial _t}{\bf{F}}\left( {{\bf{X}},t} \right)dt$ including the time component. Eventually, all the lowest order terms generated by ${\left( { - \varepsilon {L_{{\bf{g}}_1^{\bf{X}}}}} \right)^n}\left( {{{\bf{A}}_\alpha }({\bf{X}},t)\cdot d{\bf{X}}} \right)$ are in $\bf{X}$ component or time component. The terms in $\bf{X}$ component is of order $\varepsilon^{n+\alpha-(n-1)\beta}$. The terms in time component is of order $\varepsilon^{n+\alpha-(n-2)\beta-\tau}$, since time derivative can only appear one time and introduce a factor $\varepsilon^{-\tau}$.


For example, ${\left( { - \varepsilon {L_{{\bf{g}}_1^{\bf{X}}}}} \right)^2}\left( {{{\bf{A}}_\alpha }({\bf{X}},t)\cdot d{\bf{X}}} \right)$ generates a $\bf{X}$ component ${\varepsilon ^2}{\bf{g}}_1^{\bf{X}} \times {\nabla _{{{\bf{B}}_\alpha }}} \times \left( {{\bf{g}}_1^{\bf{X}} \times {{\bf{B}}_\alpha }} \right)\cdot d\mathbf{X}$ of order ${\varepsilon ^{2 + \alpha  - \beta }}$, where symbol ${\nabla _{{\alpha }}}$ denotes $\nabla$ operating on $\mathbf{B}_{\alpha}$, and time components ${\varepsilon ^2}{\bf{g}}_1^{\bf{X}}\cdot{\partial _{t,\alpha }}\left( {{\bf{g}}_1^{\bf{X}} \times {{\bf{B}}_\alpha }} \right)dt$ and $- {\varepsilon ^2}{\bf{g}}_1^{\bf{X}}\cdot{\nabla _\alpha }\left[ {{\bf{g}}_1^{\bf{X}}\cdot{\partial _t}{{\bf{A}}_\alpha }\left( {{\bf{X}},t} \right)} \right]dt$ of order ${\varepsilon ^{2 + \alpha  - \tau }}$, where symbol ${\partial _{t,\alpha }}$ denotes $\partial_t$ operating on $\mathbf{B}_\alpha$.

Then, the property of perturbation can be divided into three cases to check whether linear cancellation rule stands for each case.

\subsubsection{$\beta=1,\tau=1$}\label{sec6.9.1}

If $\beta=1,\tau=1$, which means that the spatial gradient length  and frequency of $\mathbf{B}_{\alpha}$ equals the Larmor radius and gyro frequency of charged particle, then, the lowest $\theta$-coupling terms generated by ${\left( { - \varepsilon {L_{{\bf{g}}_1^{\bf{X}}}}} \right)^n}\left( {{{\bf{A}}_\alpha }({\bf{X}},t)\cdot d{\bf{X}}} \right)$ with $n\ge 1$ are of order $\varepsilon^{1+\alpha}$. As discussed before, the terms generated at most includes time derivative only for one time. So this term can generates $n+1$ terms. One term only includes the action of $L_{\mathbf{g}_1^{\mathbf{X}}}$ as form $\mathbf{g}_1^{\mathbf{X}}\times \nabla_{\alpha}$. For the other $n$ terms, each term corresponds to the time  derivative generated by each $L_{\mathbf{g}_1^{\mathbf{X}}}$ among the $n$ production of ${\left( { - \varepsilon {L_{{\bf{g}}_1^{\bf{X}}}}} \right)^n}$.
Considering the following expanding
\begin{equation}\label{145}
{e^2} = \sum\limits_{n = 0} {\frac{{{2^n}}}{{n!}}},
\end{equation}
the series of $\sum\limits_{n = 0} {\frac{{{2^n}}}{{n!}}}$ is convergent. Therefore, the following summation
\begin{eqnarray}
\sum\limits_{n = 0} {\frac{1}{{n!}}\left| {{{\left[ {{{\left( { - \varepsilon {L_{{\bf{g}}_1^{\bf{X}}}}} \right)}^n}\left( {{{\bf{A}}_\alpha }({\bf{X}},t)\cdot d{\bf{X}}} \right)} \right]}_{1 + \alpha }}} \right|}  \le {\varepsilon ^{1 + \alpha }}\sum\limits_{n = 0} {\frac{{n + 1}}{{n!}}}
\end{eqnarray}
is a finite number, as all physical quantities are normalized.
For real calculation, we may truncate this series at a low number of $n$. And $\theta$-coupling terms included in this summation can be cancelled based on linear cancellation rule. Therefore, with condition $\beta=1,\tau=1$,  we still can carry out the cancellation with the order of $\theta$-coupling term becoming higher and higher, until at some order there is $\theta$-coupling terms in $\theta,\mu$ or $U$ components, which can not be cancelled by the linear cancellation rule.

\subsubsection{$\beta>1,\tau>1$}\label{sec6.9.2}

Term ${\left( { - \varepsilon {L_{{\bf{g}}_1^{\bf{X}}}}} \right)^n}\left( {{{\bf{A}}_\alpha }({\bf{X}},t)\cdot d{\bf{X}}} \right)$ generates $\theta$-coupling terms of order ${\varepsilon ^{n + \alpha  - (n - 1)\beta }}$ and ${\varepsilon ^{n + \alpha  - (n - 2)\beta  - \tau }}$ .
For $n$ larger , the order ${\varepsilon ^{n + \alpha  - (n - 1)\beta }}$ and ${\varepsilon ^{n + \alpha  - (n - 2)\beta  - \tau }}$ inversely becomes much lower than $\varepsilon^{1+\alpha}$. Therefore, it shouldn't simply truncate the expanding at some order, because the abandoned terms are of much lower order.

\subsubsection{$\beta<1,\tau<1$}\label{sec6.9.3}

If $\beta<1,\tau<1$, for $n$ larger , the order ${\varepsilon ^{n + \alpha  - (n - 1)\beta }}$ and ${\varepsilon ^{n + \alpha  - (n - 2)\beta  - \tau }}$ positively becomes much larger. Then, the $\theta$-coupling terms in $\mathbf{X}$ and $t$ component can be cancelled with the order of $\theta$-coupling term becoming higher and higher, until at some order there is $\theta$-coupling terms in $\theta,\mu$ or $U$ components, which can not be cancelled by the linear cancellation rule. Then, we can truncate the expanding at this order, lower than which no $\theta$-coupling term exists. And the abandoned terms are of higher order.

The analysis in Subsecs.(\ref{sec6.9.1},\ref{sec6.9.2},\ref{sec6.9.3}) can be applied to $\phi_\sigma$.

Eventually, the situation that $\theta$ can be decoupled from the remaining degrees of freedom up to some order depends on the restriction conditions imposed on perturbation that $\beta \le 1, \tau \le 1$. However, according to the calculation procedure given in Sec.(\ref{sec2}), modern GT can not give the restriction condition of perturbation.

\subsection{Approximation up to order $O(\varepsilon^2)$}\label{sec6.10}

Ref.(\cite{2016shuangxi4}) shows that the $\theta$-coupling term of $O(\varepsilon^3)$ in $\theta$ component  for guiding center motion can not be cancelled, and it derives a Lagrangian truncating the expanding up to the second order. Therefore, for the gyrocenter kinetics, our purpose is still to cancel all $\theta$-coupling terms up to order $O(\varepsilon^2)$.

So far, the introduced generators include ${\bf{g}}_1^{\bf{X}},g_1^\mu ,g_1^U,{\bf{g}}_{1 + \sigma }^{\bf{X}},g_\sigma ^\mu ,{\bf{g}}_{1 + \eta }^{\bf{X}},g_\eta ^\mu ,{\bf{g}}_{1 + \alpha }^{\bf{X}},g_\alpha ^\mu ,g_\alpha ^U,{\bf{g}}_2^{\bf{X}}, \cdots$.

If $\theta$-coupling term is in $\bf{X}$ and $t$ components, based on the linear cancellation rule, it can be canceled by introducing generators like ${\bf{g}}_{{\chi _1}}^{\bf{X}},g_{{\chi _2}}^U,g_{{\chi _3}}^\mu$ with appropriate $\chi_1,\chi_2,\chi_3$. If $\theta$-coupling terms is in one of $\theta,\mu,U$ components, they can't be cancelled based on the linear cancellation rule. Therefore, we need to find out the lowest order term which is in $\theta$, $\mu$ or $U$ components.

The original $\gamma$ given in Eq.(\ref{a11}) only includes terms in $\mathbf{X}$ and $t$ components. According to the rule given by Eq.(\ref{vp7}), the action of the operators $L_{g_{\chi}^{\mu}}$ and $L_{g_{\chi}^{U}}$ over $\mathbf{X}$ and $t$ components don't generate $(\theta,U,\mu)$ components. Only $L_{\mathbf{g}_{\chi}^{\mathbf{X}}}$ can generate $(\theta,U,\mu)$ component. The lowest linear term in $\theta$ component origins from $\varepsilon L_{\mathbf{g}_{1}^{\mathbf{X}}}\gamma_{1\mathbf{X}_{\perp}}$, the specific expression of which is ${\varepsilon ^2}{\bf{g}}_1^{\bf{X}}\cdot{\partial _\theta }\left( {{\gamma _{1{\bf{X}} \bot }}} \right)d\theta $ and equals $2{\varepsilon ^2}\mu d\theta$. The lowest-order nonlinear term in $\theta$ component origins from $\varepsilon^2 {\left( {{L_{{\bf{g}}_1^{\bf{X}}}}} \right)^2}{\gamma _0}$ and is $ - \varepsilon^2 {\bf{g}}_1^{\bf{X}}\cdot {\partial _\theta }\left( {{\bf{g}}_1^{\bf{X}} \times {{\bf{B}}_0}} \right)$ equaling $ - {\varepsilon ^2}\mu d\theta $. Therefore, there is no $\theta$-coupling term of order equal or lower than $O(\varepsilon^2)$ in $\theta, \mu$ or $U$ components.

But as Ref.(\cite{2016shuangxi4}) shows, at the order $O(\varepsilon^3)$, there is $\theta$-coupling terms in $\theta$ component. And also it can be checked that term ${L_{{\bf{g}}_1^{\bf{X}}}}{L_{g_\alpha ^\mu }}{L_{{\bf{g}}_1^{\bf{X}}}}{\gamma _0}$ generates a term ${\bf{g}}_1^{\bf{X}}\cdot{\partial _\theta }\left[ {g_\alpha ^\mu {\partial _\mu }\left( {{\bf{g}}_1^{\bf{X}} \times {\bf{B}}\left( {\bf{X}} \right)} \right)} \right]d\theta $, which is a $\theta$-coupling term in $\theta$ component and of order $\varepsilon^{2+\alpha}$. These $\theta$-coupling terms in $\theta$ component can not be cancelled based on the linear cancellation rule.

Eventually, we stop the cancellation procedure at order of $O(\varepsilon^2)$, below which no $\theta$-coupling terms exists in new coordinate system. But we also take back those lowest order terms including the effect of spatial gradient and frequency of $\mathbf{B}_{\alpha}, \nabla\phi,\partial_t \mathbf{A}_\alpha$.

\subsection{New $\Gamma$ approximated up to $O(\varepsilon^2)$ including terms of effect of spatial gradient and frequency of perturbation}

The summation of left terms in $\Gamma_0,\Gamma_{0\alpha}$,$\Gamma_{0\sigma},\Gamma_2$ and those terms including effect of spatial gradient and frequency of perturbation together leads to the approximation of $\Gamma$

\begin{equation}\label{a96}
\begin{array}{l}
\Gamma  \approx dS + \left( {{\bf{A}}\left( {\bf{X}} \right) + {{\bf{A}}_1} + \varepsilon U{\bf{b}}} \right)\cdot d{\bf{X}} + {\varepsilon ^2}\mu d\theta \\
 - \left( {\varepsilon \left( {\frac{{{U^2}}}{2} + \mu B({\bf{X}})} \right) + {\phi _\sigma}} \right)dt,
\end{array}
\end{equation}
with
\begin{eqnarray}\label{a97}
{{\bf{A}}_1} = {{\bf{A}}_\alpha }({\bf{X}},t) + {\varepsilon ^{3 + \alpha  - 2\beta }}{\left\langle {{\Gamma _{3 + \alpha  - 2\beta }}} \right\rangle _{\bf{X}}} \nonumber \\
 + {\varepsilon ^{3 + \sigma  - 2\beta }}{\left\langle {{\Gamma _{3 + \sigma  - 2\beta }}} \right\rangle _{\bf{X}}} + {\varepsilon ^{3 + \eta  - 2\beta }}{\left\langle {{\Gamma _{3 + \eta  - 2\beta }}} \right\rangle _{\bf{X}}}
\end{eqnarray}

Eq.(\ref{a97}) shows that the lowest order terms of the contribution of the spatial gradient of perturbation $\mathbf{A}_\alpha,\phi_\sigma$ to the motion of charged particles are ${\varepsilon ^{3 + \alpha  - 2\beta }}{\left\langle {{\Gamma _{3 + \alpha  - 2\beta }}} \right\rangle _{\bf{X}}}$ and ${\varepsilon ^{3 + \sigma  - 2\beta }}{\left\langle {{\Gamma _{3 + \sigma  - 2\beta }}} \right\rangle _{\bf{X}}}$, respectively, while the lowest order term of the contribution of oscillating frequency of perturbation $\mathbf{A}_\alpha$ to the motion equations is ${\varepsilon ^{3 + \eta  - 2\beta }}{\left\langle {{\Gamma _{3 + \eta  - 2\beta }}} \right\rangle _{\bf{X}}}$. These terms are obviously different from those terms derived by modern GT. The reason for the difference is that modern GT doesn't adopt a correct method to systematically decouple gyroangle from the rest degrees of freedom.  Another obvious difference is a that there is a so-called finite-radius-effect term proportional to $\left\langle {{{\left( {\bm{\rho} \cdot\nabla } \right)}^2}{\phi _\sigma }} \right\rangle dt$ in the eventually approximated Lagrangian 1-form derived by modern GT. However, according to our theory, Eq.(\ref{a74}) shows that ${\varepsilon ^{ - \sigma }}{\bf{g}}_1^{\bf{X}}\cdot\nabla {\phi _\sigma }dt$ doesn't appear independently, but is cancelled by another term $g_\sigma ^\mu {\partial _\mu }{\gamma _{1t}}dt$, and all these terms form a gauge term $\Gamma_{1+\sigma}$ totally as shown in Eq.(\ref{a74}), which doesn't contribute to the motion of charged particle. Therefore, at the next order, term ${\varepsilon ^{ - \sigma }}{L_{{\bf{g}}_1^{\bf{X}}}}\left( {{\bf{g}}_1^{\bf{X}}\cdot\nabla \phi dt} \right)$, which can produce the finite radius effect term $\left\langle {{{\left( {\bm{\rho} \cdot\nabla } \right)}^2}{\phi _\sigma }} \right\rangle dt$, doesn't appear independently, but is included by a gauge term ${L_{{\bf{g}}_1^{\bf{X}}}}\left( {{\Gamma _{1 + \sigma }}} \right)$, from which the finite radius effect term can't not be extracted out.

However, the terms given by Eqs.(\ref{a150},\ref{a157},\ref{a166}) in \ref{app2} are complex to remove sine and cosine functions of $\theta$ from them. One reason is that in this paper, it's assumed that the wave length of perturbation in each direction is of the same order that all directions take part in the calculation at the same order. However, for real physical configuration, the wave length of perturbation for each direction present disparate scales. For example, for plasma in tokamak, the radial wave length is much larger than that in poloidal and toroidal direction. It's expected that this kind of difference could simplify the removing calculation much. But we will carry out the work concerned with disparate length scale of wave length in each direction in another paper.

The motion equations are derived based on Lagrangian 1-form given by Eq.(\ref{a96})
\begin{equation}\label{a32}
\mathop {\bf{X}}\limits^. {\rm{ = }}\frac{{U{{\bf{B}}} + {\bf{b}} \times \nabla H + \partial {{\bf{A}}_1}/\partial t \times {\bf{b}}}}{{{\bf{b}}\cdot{{\bf{B}}^*}}},
\end{equation}
\begin{eqnarray}\label{b19}
\dot U =&& \frac{{ - \left( {{{\bf{B}}^*}\cdot\nabla H} \right) + {\bf{b}}\cdot\left[ {\left( {\partial {{\bf{A}}_1}/\partial t \times {\bf{b}}} \right) \times {{\bf{B}}^*}} \right]}}{{{\varepsilon\bf{b}}\cdot{{\bf{B}}^*}}}\nonumber \\
&& - {\bf{b}}\cdot\frac{{\partial {{\bf{A}}_1 }}}{{\varepsilon\partial t}},
\end{eqnarray}
with ${{\bf{B}}^*} = \nabla  \times \left( {{\bf{A}}\left( {\bf{X}} \right) + {{\bf{A}}_1} + \varepsilon U{\bf{b}}} \right)$ and $H = \varepsilon \left( {\frac{{{U^2}}}{2} + \mu B({\bf{X}}) + {\phi _1}} \right)$.
Eq.(\ref{b19}) shows the acceleration of parallel velocity by inductive electric field.

In the motion equations Eqs,(\ref{a32},\ref{b19}), the terms related with perturbation are ${{\bf{B}}_\alpha }\left( {{\bf{X}},t} \right),\partial {{\bf{A}}_\alpha }\left( {{\bf{X}},t} \right)/\partial t,\nabla {\phi _\sigma }$ and the action of operator over these three terms, which are physical quantities.

%

\section{Solving the coordinate transform.}\label{sec8}

Based on the cancellation procedure in Sec.(\ref{sec5}), there are six independent parameters  $\varepsilon,\varepsilon^{\alpha},\varepsilon^{\eta},\varepsilon^{\sigma},\varepsilon^{-\beta},\varepsilon^{-\tau}$, which will be renamed to be $(\varepsilon_1,\varepsilon_2,\varepsilon_3,\varepsilon_4,\varepsilon_5,\varepsilon_6)$, respectively. With these six basic parameters, the old coordinate can be formulated to be
\begin{equation}\label{a101}
Z_b^\nu ({\bf{Z}},{\bf{E}}_6) = {Z^\nu } + Z_b^{\nu *}({\bf{Z}},{\bf{E}}_6),
\end{equation}
with
\begin{equation}\label{a102}
Z_b^{\nu *}({\bf{Z}},{{\bf{E}}_6}) = \sum\limits_{{m_1} \ge 0, \cdots ,{m_6} \ge 0}^* {\varepsilon _1^{{m_1}} \cdots \varepsilon _6^{{m_5}}Z_{b,{m_1}{m_2}{m_3}{m_4}{m_5}{m_6}}^\nu ({\bf{Z}})}
\end{equation}
where $*$ means that term with $m_1=m_2=m_3=m_4=m_5=m_6=0$ is deleted. The expanding of $\gamma_{(0,0\alpha,0\sigma,1)}$ are
\begin{equation}\label{a103}
\begin{array}{*{20}{l}}
{{{\bf{\gamma }}_{\left( {0,\sigma ,\alpha ,1} \right)}} = \sum\limits_{n = 0} {\frac{1}{{n!}}{{\left( {{\bf{Z}}_b^*} \right)}^n}:\nabla _{\bf{Z}}^n{\gamma _{\left( {0,\sigma ,\alpha ,1} \right)}}\left( {\bf{Z}} \right)} }\\
{ = {\gamma _{\left( {0,\sigma ,\alpha ,1} \right)}}\left( {\bf{Z}} \right) + {\varepsilon _1}{{\bf{Z}}_{b,100000}}\cdot{\nabla _{\bf{Z}}}{\gamma _{\left( {0,\sigma ,\alpha ,1} \right)}} + {\varepsilon _2}{{\bf{Z}}_{b,010000}}\cdot{\nabla _{\bf{Z}}}{\gamma _{\left( {0,\sigma ,\alpha ,1} \right)}}}\\
{ + {\varepsilon _3}{{\bf{Z}}_{b,001000}}\cdot{\nabla _{\bf{Z}}}{\gamma _{\left( {0,\sigma ,\alpha ,1} \right)}} + {\varepsilon _4}{{\bf{Z}}_{b,000100}}\cdot{\nabla _{\bf{Z}}}{\gamma _{\left( {0,\sigma ,\alpha ,1} \right)}}}\\
{ + {\varepsilon _5}{{\bf{Z}}_{b,000010}}\cdot{\nabla _{\bf{Z}}}{\gamma _{\left( {0,\sigma ,\alpha ,1} \right)}} + {\varepsilon _6}{{\bf{Z}}_{b,000001}}\cdot{\nabla _{\bf{Z}}}{\gamma _{\left( {0,\sigma ,\alpha ,1} \right)}} + ( \cdots ).}
\end{array}
\end{equation}
where each subscript in $\gamma_{({\left( {0,\sigma ,\alpha ,1} \right)})}$ represents each kind of 1-form, e.g,$\sigma$ represents $\gamma_{\sigma}$.

The pullback transform formula of $\gamma(\mathbf{Z}_b(\mathbf{Z},\mathbf{E}_6))$ given by Eq.(\ref{a11}) is given by Eq.(\ref{a2})
\begin{equation}\label{a104}
{\Gamma _\nu }\left( {{\bf{Z}},\mathbf{E}_6 } \right) = \frac{{\partial Z_b^v\left( {{\bf{Z}},{{\bf{E}}_6}} \right)}}{{\partial Z_{}^\nu }}\left( {{\gamma _0} + {\gamma _{\alpha }} + {\gamma _{\sigma }} + \varepsilon_1 {\gamma _1}} \right)\left( {{{\bf{Z}}_b}\left( {{\bf{Z}},{{\bf{E}}_6}} \right)} \right).
\end{equation}
$\Gamma$ on LHS of Eq.(\ref{a104}) is already derived and  given by Eq.(\ref{a96}). The RHS of Eq.(\ref{a104}) is recorded as $\bar{\Gamma}$. It's found that ${\Gamma _0} = {{\bar \Gamma }_0} = {\bf{A}}\left( {\bf{X}} \right)\cdot d{\bf{X}}$, ${\Gamma _{\alpha }} = {{\bar \Gamma }_{\alpha }} = {{\bf{A}}_\alpha }\left( {{\bf{X}},t} \right)\cdot d{\bf{X}}$, ${\Gamma _{\sigma }} = {{\bar \Gamma }_{\sigma }} = {\phi _\sigma }\left( {{\bf{X}},t} \right)dt$, which are trival to solve $\mathbf{Z}_b$.

\subsection{The order of $O(\varepsilon_1)$}

For the order of $O(\varepsilon_1)$, $\Gamma_{100000}$ is given as
\begin{equation}\label{a105}
{\Gamma _{100000}} = U{\bf{b}}\cdot d{\bf{X}} - \left( {\frac{{{U^2}}}{2} + \mu B\left( {\bf{X}} \right)} \right)dt.
\end{equation}
The subscripts of $\Gamma$ are of the same meaning given in Eq.(\ref{a102}), and are also applied to $\bar{\Gamma}$. $\bar{\Gamma}_{100000,\mathbf{X}}$ is
\begin{equation}\label{a106}
\begin{array}{*{20}{l}}
{{{\bar \Gamma }_{100000,{\bf{X}}}} = \frac{{\partial Z_{b,100000}^\mu }}{{\partial {\bf{X}}}}{A_\mu } + Z_{b,100000}^\mu {\partial _\mu }{\bf{A}} + U{\bf{b}} + \sqrt {2B\left( {\bf{X}} \right)\mu } {{\widehat {\bf{v}}}_ \bot }{\rm{ }}}\\
{ =  - {\bf{Z}}_{b,100000}^{\bf{X}} \times {\bf{B}}\left( {\bf{X}} \right) + U{\bf{b}} + \sqrt {2B\left( {\bf{X}} \right)\mu } {{\widehat {\bf{v}}}_ \bot } + \nabla \left( {Z_{b,100000}^\mu {A_\mu }} \right)}
\end{array}
\end{equation}
and
\begin{equation}\label{a107}
{{\bar \Gamma }_{100000,(t,\mu ,\theta ,\mu )}} = \frac{{\partial Z_{b,100000}^\mu }}{{\partial (t,\mu ,\theta ,\mu )}}{A_\mu } = \frac{\partial }{{\partial (t,\mu ,\theta ,\mu )}}\left( {Z_{b,100000}^\mu {A_\mu }} \right).
\end{equation}
Eqs.(\ref{a106},\ref{a107}) show that $\left( {Z_{b,100000}^\mu {A_\mu }} \right)$ is a gauge term, and the solution of ${\bf{Z}}_{b,100000}^{\bf{X}}$ is
\begin{equation}\label{a108}
{\bf{Z}}_{b,100000}^{\bf{X}} = \bm{\rho}.
\end{equation}

\subsection{The order of $O(\varepsilon_1\varepsilon_2)$}

For the order of $O(\varepsilon_1\varepsilon_2)$, it's derived that
\begin{equation}\label{a1108}
{\Gamma _{110000}} = 0,
\end{equation}
which leads to the following identity
\begin{equation}\label{a109}
\begin{array}{*{20}{l}}
{{{\bar \Gamma }_{110000,v}} = \frac{1}{2}{{\bf{Z}}_{b,100000}}{{\bf{Z}}_{b,010000}}:\nabla _{\bf{Z}}^2{A_v} + \frac{{\partial Z_{b,100000}^\mu }}{{\partial {Z^v}}}\left( {{{\bf{Z}}_{b,010000}}\cdot\nabla _{\bf{Z}}^{}} \right){A_\mu }}\\
{ + \frac{{\partial Z_{b,010000}^\mu }}{{\partial {Z^v}}}{{\bf{Z}}_{b,100000}}\cdot\nabla _{\bf{Z}}^{}{A_\mu } + {{\bf{Z}}_{b,110000}}\cdot\nabla _{\bf{Z}}^{}{A_v} + \frac{{\partial Z_{b,110000}^\mu }}{{\partial {Z^v}}}{A_\mu }{\rm{ }}}\\
{ + \left( {{{\bf{Z}}_{b,100000}}\cdot\nabla _{\bf{Z}}^{}} \right){A_{\alpha v}} + \frac{{\partial Z_{b,100000}^\mu }}{{\partial {Z^v}}}{A_{\alpha \mu }} + {{\bf{Z}}_{b,010000}}\cdot\nabla _{\bf{Z}}^{}{\gamma _{1v}} = 0}
\end{array}
\end{equation}
For $v\in(X^1,X^2,X^3)$, it's derived that
\begin{equation}\label{a110}
- {\bf{Z}}_{b,110000}^{\bf{X}} \times {\bf{B}}\left( {\bf{X}} \right) - {\bf{Z}}_{b,100000}^{\bf{X}} \times \left( {\nabla  \times {{\bf{A}}_\alpha }} \right) + {{\bf{Z}}_{b,010000}}\cdot\nabla _{\bf{Z}}^{}{\gamma _{1{\bf{X}}}} = 0.
\end{equation}
For $v=t$, it's derived that
\begin{equation}\label{a111}
{{\bf{Z}}_{b,010000}}\cdot\nabla _{\bf{Z}}^{}{\gamma _{1t}} = 0.
\end{equation}
Eqs.(\ref{a110},\ref{a111}) give the solutions of ${\bf{Z}}_{b,110000}^{\bf{X}},Z_{b,010000}^U,Z_{b,010000}^\mu $
\begin{equation}\label{a112}
{\bf{Z}}_{b,110000}^{\bf{X}} =  - \frac{{\bf{b}}}{{B\left( {\bf{X}} \right)}} \times {\left[ {\bm{\rho}  \times {{\bf{B}}_\alpha } - \left( {Z_{b,010000}^U{\partial _U} + Z_{b,010000}^\mu {\partial _\mu }} \right){\gamma _{1{\bf{X}} \bot }}} \right]_ \bot },
\end{equation}
\begin{equation}\label{a113}
Z_{b,010000}^U = {\left( {\bm{\rho}  \times {{\bf{B}}_\alpha }} \right)_\parallel }/{\partial _U}{\gamma _{1{\bf{X}}\parallel }},
\end{equation}
\begin{equation}\label{a114}
Z_{b,010000}^\mu  =  - UZ_{b,010000}^U/B\left( {\bf{X}} \right).
\end{equation}

The higher order terms can be solved in the same way.

\section{Summary and Discussion}\label{sec7}

In this paper, it's pointed out that the single-parameter LTPT can not be applied straightforwardly to decouple gyroangle from the remaining degrees of freedom in the Lagrangian 1-form of a charged particle in the magnetized plasma, where the perturbations present multiple scales. The application of the single-parameter LTPT by modern GT to this Lagrangian 1-form leads to two issues. one is the confusion of the order of those small perturbation parameters, which results in inappropriate amplification of the generators and leads to the violation of NIT by the coordinate transform. The other one is that nonphysical terms appears in the trajectory equations.

To overcome these two issues, instead of the single-parameter LPTP, we utilize a new multi-parameter Lie  transform method which is first introduced in Ref.(\cite{2016shuangxi4}). The application of this method must be assisted by a kind of linear cancellation rule. By applying this method to the Lagrangian 1-form, new generators need to be introduced to the exponential operator $\exp \left( { - {\bf{E}}\cdot{L_{\bf{g}}}} \right)$ order by order. The new introduced generators  generate terms, the lowest order ones of which are used to cancel the existed $\theta$-coupling terms, and the left $\theta$-coupling terms are of higher order than that of the cancelled terms.

This procedure of cancelling $\theta$-coupling terms can be carried out to $O(\varepsilon^2)$. Furthermore, the eventual approximate Lagrangian 1-form given in Eq.(\ref{a96}) presents terms accounting for the effect of spatial gradient and oscillating frequency of the perturbations different from those derived by modern GT.


\section{Acknowledgement}

This work was completed at Uji Campus, Kyoto University, Japan. The author thanks the communication with Prof. Yasuaki Kishimoto, Prof. Jiquan Li, Prof. Kenji Imadera,  Prof. Zhiyong Qiu, Dr.Zhixi.Lu, Dr.Defeng Kong, Dr. Zanhui Wang and Dr.Ming Xu.

\appendix

\section{Simple introduction to John Cary's LPTT \cite{1983cary}adopted by modern GT}\label{app1}

This theory begins with the following autonomous differential equations
\begin{equation}\label{a117}
\frac{{\partial Z_f^\mu }}{{\partial \epsilon }}\left( {{\bf{z}},\epsilon } \right) = {g_1^\mu }\left( {{{\bf{Z}}_f}\left( {{\bf{z}},\epsilon } \right)} \right),
\end{equation}
\begin{equation}\label{a118}
\frac{{d{\bf{z}}}}{{d\varepsilon }} = 0,
\end{equation}
where $\mathbf{Z}=\mathbf{Z}_f(\mathbf{z},\epsilon)$ is new coordinate, $\mathbf{z}$ is old coordinate, and $\epsilon$ is an independent variable denoting the small parameter of amplitude of perturbation.
Eqs.(\ref{a117}) and (\ref{a118}) lead to the solution
\begin{equation}\label{c1}
{\bf{z}} = \exp \left(-{{\epsilon g^i_1}{\partial _{{Z_i}}}} \right){\bf{Z}}.
\end{equation}
For a differential 1-form written as $\gamma(\bf{z})$, which doesn't depend on $\epsilon$ in the coordinate frame of $\bf{z}$, coordinate transform given by Eq.(\ref{c1}) induces a pullback transform of $\gamma$ as
\begin{equation}\label{c2}
{\Gamma _\mu }\left(\mathbf{ Z} \right) = {\left[ {\exp \left( { - \varepsilon {L_{1}}} \right)\gamma } \right]_\mu }\left( \mathbf{Z} \right) + \frac{{\partial S\left( \mathbf{Z} \right)}}{{\partial {Z^\mu }}}dZ^\mu.
\end{equation}
where $S(\mathbf{Z})$ is a gauge function and the $\mu$ component of $L_1\gamma$ is defined as ${\left( {{L_1}\gamma } \right)_\mu } = g_1^i\left( {{\partial _i}{\gamma _\mu } - {\partial _\mu }{\gamma _i}} \right)$.

When differential 1-form explicitly depends on the perturbation and can be written as
$\gamma(\mathbf{z},\varepsilon)  = {\gamma _0}(\mathbf{z}) + \epsilon {\gamma _1}(\mathbf{z}) + \epsilon ^{2}{\gamma _2}(\mathbf{z}) +  \cdots$,
Ref.(\cite{1983cary}) generalize Eq.(\ref{c2}) to be a composition of individual Lie transforms $T =  \cdots {T_3}{T_2}{T_1}$ with
\begin{equation}\label{c4}
{T_n} = \exp \left( { - \epsilon^n {L_{n}}} \right),
\end{equation}
to get the new 1-form
\begin{equation}\label{f1}
\Gamma  = T\gamma  + dS,
\end{equation}
which can be expanded by the order of $\epsilon$
\begin{equation}\label{c5}
{\Gamma _0} = {\gamma _0},
\end{equation}
\begin{equation}\label{c6}
{\Gamma _1} = d{S_1} - {L_1}{\gamma _0} + {\gamma _1},
\end{equation}
\begin{equation}\label{c7}
{\Gamma _2} = d{S_2} - {L_2}{\gamma _0} + {\gamma _2} - {L_1}{\gamma _1} + \frac{1}{2}L_1^2{\gamma _0},
\end{equation}
\begin{equation}\label{a33}
\cdots   \nonumber \\
\end{equation}
There expansion formulas can be written in a general form
\begin{equation}\label{c8}
\Gamma_n = d S_n - L_n \gamma_0 + C_n.
\end{equation}
By requiring $\Gamma_{ni}=0,i\in(1,\cdots,2N)$, the $n$th order generators are
\begin{equation}\label{c9}
g_n^j = \left( {\frac{{\partial {S_n}}}{{\partial {z^i}}} + {C_{ni}}} \right)J_0^{ij},
\end{equation}
where $J_0^{ij}$ is Poisson tensor.
And correspondingly, the $n$th order gauge function can be solved as
\begin{equation}\label{c10}
V_0^\mu \frac{{\partial {S_n}}}{{\partial {z^\mu }}} = \frac{{\partial {S_n}}}{{\partial {z^0}}} + V_0^i\frac{{\partial {S_n}}}{{\partial {z^i}}} = {\Gamma _{n0}} - {C_{n\mu }}V_0^\mu
\end{equation}
with
\begin{equation}\label{c11}
V_0^i = J_0^{ij}\left( {\frac{{\partial {\gamma _{0j}}}}{{\partial {z^0}}} - \frac{{\partial {\gamma _{00}}}}{{\partial {z^j}}}} \right)
\end{equation}
To avoid the secularity of $S_n$, usually $\Gamma_{n0}$ is chosen to be
\begin{equation}\label{c12}
{\Gamma _{n0}} = \left[\kern-0.15em\left[ {V_0^\mu {C_{n\mu }}}
 \right]\kern-0.15em\right],
\end{equation}
where $\left[\kern-0.15em\left[  \cdots
 \right]\kern-0.15em\right]$ means average over the fast variable.

\section{Calculating the terms of effect including the effect of spatial gradient length and frequency of perturbation}\label{app2}

\subsection{Terms including effect of spatial gradient and frequency of $\mathbf{B}_{\alpha}(\mathbf{X},t)$}

The terms including effect of spatial gradient and frequency of $\mathbf{B}_{\alpha}(\mathbf{X},t)$ are given in Eqs.(\ref{a130}-\ref{a132}). It's easy to notice that ${\left( {{L_{{\bf{g}}_1^{\bf{X}}}}{L_{{\bf{g}}_1^{\bf{X}}}}{L_{g_\alpha ^U}}{L_{{\bf{g}}_1^{\bf{X}}}}{\gamma _0}} \right)_{3 + \alpha  - 2\beta }}=0$, since ${L_{g_\alpha ^U}}{L_{{\bf{g}}_1^{\bf{X}}}}{\gamma _0} =  - {L_{g_\alpha ^U}}\left( {{\bf{g}}_1^{\bf{X}} \times {\bf{B}}\left( {\bf{X}} \right)\cdot d{\bf{X}}} \right)=0$. ${\bf{g}}_1^{\bf{X}}$ and $g_\alpha ^\mu$ are given in Eqs.(\ref{a22}) and (\ref{a83}), respectively. It's derived that
\begin{equation}\label{a145}
{L_{{\bf{g}}_1^{\bf{X}}}}\left( {{\bf{A}}\cdot d{\bf{X}}} \right) = \sqrt {2\mu B\left( {\bf{X}} \right)} {\widehat {\bf{v}}_ \bot }\cdot d{\bf{X}},
\end{equation}
\begin{equation}\label{a146}
{L_{g_\alpha ^\mu }}{L_{{\bf{g}}_1^{\bf{X}}}}{\gamma _0} = g_\alpha ^\mu {\partial _\mu }\left( {{L_{{\bf{g}}_1^{\bf{X}}}}{\gamma _0}} \right) = \frac{{{\varepsilon ^{ - \alpha }}U}}{{ B\left( {\bf{X}} \right)}}\left( {{{\widehat {\bf{v}}}_ \bot }\cdot{{\bf{B}}_\alpha }} \right){\widehat {\bf{v}}_ \bot }\cdot d\bf{X}.
\end{equation}
With the following vector identities
\begin{equation}\label{a147}
{\nabla _{{{\bf{B}}_\alpha }}} \times \left( {\left( {{{\widehat {\bf{v}}}_ \bot }\cdot{{\bf{B}}_\alpha }} \right){{\widehat {\bf{v}}}_ \bot }} \right) = {\nabla _{{{\bf{B}}_\alpha }}}\left( {{{\widehat {\bf{v}}}_ \bot }\cdot{{\bf{B}}_\alpha }} \right) \times {\widehat {\bf{v}}_ \bot },
\end{equation}
where ${\nabla _{{{\bf{B}}_\alpha }}}$ means that $\nabla$ acts on $\mathbf{B}_\alpha$, and
\begin{equation}\label{a148}
\hat {\bm{\rho}}  \times [{\nabla _{{{\bf{B}}_\alpha }}} \times \left( {\left( {{{\widehat {\bf{v}}}_ \bot }\cdot{{\bf{B}}_\alpha }} \right){{\widehat {\bf{v}}}_ \bot }} \right)] =  - {\widehat {\bf{v}}_ \bot }\cdot\left( {\hat {\bm{\rho}} \cdot\nabla {{\bf{B}}_\alpha }} \right){\widehat {\bf{v}}_ \bot },
\end{equation}
\begin{equation}\label{a149}
\hat{\bm{\rho}}  \times [{\nabla _{{{\bf{B}}_\alpha }}} \times \left( { - {{\widehat {\bf{v}}}_ \bot }\cdot\left( {\hat{\bm{\rho}} \cdot\nabla {{\bf{B}}_\alpha }} \right){{\widehat {\bf{v}}}_ \bot }} \right)] = {\widehat {\bf{v}}_ \bot }\cdot\left[ {{{\left( {\hat {\bm{\rho}} \cdot\nabla } \right)}^2}{{\bf{B}}_\alpha }} \right]{\widehat {\bf{v}}_ \bot },
\end{equation}
where vector identity ${\bf{A}} \times \left( {{\bf{B}} \times {\bf{C}}} \right) = \left( {{\bf{A}}\cdot{\bf{C}}} \right){\bf{B}} - \left( {{\bf{A}}\cdot{\bf{B}}} \right){\bf{C}}$ is used,
it's derived that
\begin{equation}\label{a150}
\begin{array}{l}
{\left( {{L_{{\bf{g}}_1^{\bf{X}}}}{L_{{\bf{g}}_1^{\bf{X}}}}{L_{g_\alpha ^\mu }}{L_{{\bf{g}}_1^{\bf{X}}}}{\gamma _0}} \right)_{3 + \alpha  - 2\beta }}\\
 = \frac{{{\varepsilon ^{ - \alpha }}\mu U}}{{{B^2}}}\hat {\bm{\rho}}  \times \left[ {{\nabla _{{{\bf{B}}_\alpha }}} \times \left( {\hat{\bm{\rho}}  \times \left( {{\nabla _{{{\bf{B}}_\alpha }}} \times \left( {\left( {{{\widehat {\bf{v}}}_ \bot }\cdot{{\bf{B}}_\alpha }} \right){{\widehat {\bf{v}}}_ \bot }} \right)} \right)} \right)} \right]\cdot d{\bf{X}}\\
 = \frac{{{\varepsilon ^{ - \alpha }}\mu U}}{{ {B^2}}}{\widehat {\bf{v}}_ \bot }\cdot\left[ {{{\left( {\hat{\bm{\rho}} \cdot\nabla } \right)}^2}{{\bf{B}}_\alpha }} \right]{\widehat {\bf{v}}_ \bot }\cdot d{\bf{X}}.
\end{array}
\end{equation}
Therefore, the effective contribution of effect of spatial gradient of $\mathbf{B}_\alpha$ is
\begin{equation}\label{a151}
\begin{array}{l}
{\varepsilon ^{3 + \alpha  - 2\beta }}{\left\langle {{\Gamma _{3 + \alpha  - 2\beta }}} \right\rangle _{\bf{X}}} = \frac{1}{{12}}{\varepsilon ^{3 + \alpha }}\left\langle {{{\left( {{L_{{\bf{g}}_1^{\bf{X}}}}{L_{{\bf{g}}_1^{\bf{X}}}}{L_{g_\alpha ^\mu }}{L_{{\bf{g}}_1^{\bf{X}}}}{\gamma _0}} \right)}_{3 + \alpha  - 2\beta }}} \right\rangle \\
 = \frac{1}{{12}}{\varepsilon ^3}\frac{{\mu U}}{{{B^2}}}\left\langle {{{\widehat {\bf{v}}}_ \bot }\cdot\left[ {{{\left( {\hat {\bm{\rho}} \cdot\nabla } \right)}^2}{{\bf{B}}_\alpha }} \right]{{\widehat {\bf{v}}}_ \bot }} \right\rangle
\end{array}
\end{equation}
where $\left\langle {} \right\rangle $ denotes that sine and cosine functions are removed. It can also be derived that
\begin{equation}\label{a152}
\begin{array}{l}
{\left( {{L_{{\bf{g}}_1^{\bf{X}}}}{L_{{\bf{g}}_1^{\bf{X}}}}{L_{g_\alpha ^\mu }}{L_{{\bf{g}}_1^{\bf{X}}}}{\gamma _0}} \right)_{3 + \alpha  - \beta  - \tau }}\\
 = \frac{{{\varepsilon ^{ - \alpha }}\mu U}}{{ {B^2}}}\left[ \begin{array}{l}
\hat \rho \cdot{\nabla _{{{\bf{B}}_\alpha }}}\left( {\hat {\bm{\rho}} \cdot{\partial _t}\left( {\left( {{{\widehat {\bf{v}}}_ \bot }\cdot{{\bf{B}}_\alpha }} \right){{\widehat {\bf{v}}}_ \bot }} \right)} \right)\\
 + \hat{\bm{\rho}} \cdot{\partial _t}\left( {{{\widehat {\bf{v}}}_ \bot }\cdot\left( {\hat{\bm{\rho}} \cdot\nabla {{\bf{B}}_\alpha }} \right){{\widehat {\bf{v}}}_ \bot }} \right)
\end{array} \right]dt = 0
\end{array}
\end{equation}
where Eq.(\ref{a148}) and identity $\hat{\rho}\cdot \hat{\mathbf{v}}_\perp=0$ are used.

\subsection{Terms including effect of spatial gradient and frequency of $\nabla \phi_\sigma$}

The terms including effect of spatial gradient and frequency of $\mathbf{B}_{\alpha}(\mathbf{X},t)$ are given in Eqs.(\ref{a136}-\ref{a138}). $g_\sigma ^\mu $ is given in Eq.(\ref{a78}). With the following vector identities
\begin{equation}\label{a153}
{L_{g_\sigma ^\mu }}{L_{{\bf{g}}_1^{\bf{X}}}}{\gamma _0} = \frac{{{\varepsilon ^{ - \sigma }}\hat{\bm{\rho}} \cdot\nabla {\phi _\sigma }}}{{B\left( {\bf{X}} \right)}}{\widehat {\bf{v}}_ \bot }\cdot d{\bf{X}},
\end{equation}
\begin{equation}\label{a154}
{\nabla _{{\phi _\sigma }}} \times \left( {\left( {\hat{\bm{\rho}} \cdot\nabla {\phi _\sigma }} \right){{\widehat {\bf{v}}}_ \bot }} \right) = {\nabla _{{\phi _\sigma }}}\left( {\hat{\bm{\rho}} \cdot\nabla {\phi _\sigma }} \right) \times {\widehat {\bf{v}}_ \bot },
\end{equation}
\begin{equation}\label{a155}
\hat{\bm{\rho}}  \times [{\nabla _{{\phi _\sigma }}} \times \left( {\left( {\hat{\bm{\rho}} \cdot\nabla {\phi _\sigma }} \right){{\widehat {\bf{v}}}_ \bot }} \right)] =  - \left( {{{\left( {\hat{\bm{\rho}} \cdot\nabla } \right)}^2}{\phi _\sigma }} \right){\widehat {\bf{v}}_ \bot },
\end{equation}
\begin{equation}\label{a156}
\hat{\bm{\rho}}  \times [{\nabla _{{\phi _\sigma }}} \times \left( { - \left( {{{\left( {\hat{\bm{\rho}} \cdot\nabla } \right)}^2}{\phi _\sigma }} \right){{\widehat {\bf{v}}}_ \bot }} \right)] = \left[ {{{\left( {\hat{\bm{\rho}} \cdot\nabla } \right)}^3}{\phi _\sigma }} \right]{\widehat {\bf{v}}_ \bot },
\end{equation}
it can be derived that
\begin{equation}\label{a157}
\begin{array}{l}
{\left( {{L_{{\bf{g}}_1^{\bf{X}}}}{L_{{\bf{g}}_1^{\bf{X}}}}{L_{g_\sigma ^\mu }}{L_{{\bf{g}}_1^{\bf{X}}}}{\gamma _0}} \right)_{3 + \alpha  - 2\beta }}\\
 = \frac{{2\mu {\varepsilon ^{ - \sigma }}}}{{B{{\left( {\bf{X}} \right)}^2}}}\hat{\bm{\rho}}  \times \left[ {{\nabla _{{\phi _\sigma }}} \times \left( {\hat{\bm{\rho}}  \times \left( {{\nabla _{{\phi _\sigma }}} \times \left( {\left( {\hat{\bm{\rho}} \cdot\nabla {\phi _\sigma }} \right){{\widehat {\bf{v}}}_ \bot }} \right)} \right)} \right)} \right]\cdot d{\bf{X}}\\
 = \frac{{2\mu {\varepsilon ^{ - \sigma }}}}{{B{{\left( {\bf{X}} \right)}^2}}}\left[ {{{\left( {\hat{\bm{\rho}} \cdot\nabla } \right)}^3}{\phi _\sigma }} \right]{\widehat {\bf{v}}_ \bot }.
\end{array}
\end{equation}
Therefore, the effective contribution of effect of spatial gradient of $\nabla \phi_\sigma$ is
\begin{equation}\label{158}
\begin{array}{l}
{\varepsilon ^{3 + \sigma  - 2\beta }}{\left\langle {{\Gamma _{3 + \sigma  - 2\beta }}} \right\rangle _{\bf{X}}} = \frac{1}{{24}}{\varepsilon ^{3 + \sigma }}\left\langle {{{\left( {{L_{{\bf{g}}_1^{\bf{X}}}}{L_{{\bf{g}}_1^{\bf{X}}}}{L_{g_\sigma ^\mu }}{L_{{\bf{g}}_1^{\bf{X}}}}{\gamma _0}} \right)}_{3 + \sigma  - 2\beta }}} \right\rangle \\
 = \frac{1}{{12}}\frac{{{\varepsilon ^3}\mu }}{{B{{\left( {\bf{X}} \right)}^2}}}\left\langle {\left[ {{{\left( {\hat{\bm{\rho}} \cdot\nabla } \right)}^3}{\phi _\sigma }} \right]{{\widehat {\bf{v}}}_ \bot }} \right\rangle
\end{array}
\end{equation}

Similar to Eq.(\ref{a152}), it can be derived that
\begin{equation}\label{a159}
{\left( {{L_{{\bf{g}}_1^{\bf{X}}}}{L_{{\bf{g}}_1^{\bf{X}}}}{L_{g_\sigma ^\mu }}{L_{{\bf{g}}_1^{\bf{X}}}}{\gamma _0}} \right)_{3 + \sigma  - \beta  - \tau }} = 0.
\end{equation}

\subsection{Terms including effect of spatial gradient and frequency of $\partial_t \mathbf{A}_\alpha$}

The terms including effect of spatial gradient and frequency of $\partial_t \mathbf{A}_\alpha$ are given in Eqs.(\ref{a141}-\ref{a143}). $g_\eta ^\mu $ is given in Eq.(\ref{a78}). With the following vector identities
\begin{equation}\label{a162}
{L_{g_\eta ^\mu }}{L_{{\bf{g}}_1^{\bf{X}}}}{\gamma _0} = \frac{{{\varepsilon ^{ - \eta }}\hat{\bm{\rho}} \cdot{\partial _t}{{\bf{A}}_\alpha }}}{{B\left( {\bf{X}} \right)}}{\widehat {\bf{v}}_ \bot }\cdot d{\bf{X}},
\end{equation}
\begin{equation}\label{a163}
{\nabla _{{{\bf{A}}_\alpha }}} \times \left( {\left( {\hat{\bm{\rho}} \cdot{\partial _t}{{\bf{A}}_\alpha }} \right){{\widehat {\bf{v}}}_ \bot }} \right) = {\nabla _{{{\bf{A}}_\alpha }}}\left( {\hat{\bm{\rho}} \cdot{\partial _t}{{\bf{A}}_\alpha }} \right) \times {\widehat {\bf{v}}_ \bot },
\end{equation}
\begin{equation}\label{a164}
\hat{\bm{\rho}}  \times [{\nabla _{{{\bf{A}}_\alpha }}} \times \left( {\left( {\hat{\bm{\rho}} \cdot{\partial _t}{{\bf{A}}_\alpha }} \right){{\widehat {\bf{v}}}_ \bot }} \right)] =  - \hat{\bm{\rho}} \cdot\left( {\left( {\hat{\bm{\rho}} \cdot\nabla } \right){\partial _t}{{\bf{A}}_\alpha }} \right){\widehat {\bf{v}}_ \bot },
\end{equation}
\begin{equation}\label{a165}
\hat{\bm{\rho}}  \times [{\nabla _{{{\bf{A}}_\alpha }}} \times \left( { - \hat{\bm{\rho}} \cdot\left( {\left( {\hat{\bm{\rho}} \cdot\nabla } \right){\partial _t}{{\bf{A}}_\alpha }} \right){{\widehat {\bf{v}}}_ \bot }} \right)] = \hat{\bm{\rho}} \cdot\left[ {{{\left( {\hat{\bm{\rho}} \cdot\nabla } \right)}^2}{\partial _t}{{\bf{A}}_\alpha }} \right]{\widehat {\bf{v}}_ \bot },
\end{equation}
it can be derived that
\begin{equation}\label{a166}
\begin{array}{l}
{\left( {{L_{{\bf{g}}_1^{\bf{X}}}}{L_{{\bf{g}}_1^{\bf{X}}}}{L_{g_\eta ^\mu }}{L_{{\bf{g}}_1^{\bf{X}}}}{\gamma _0}} \right)_{3 + \eta  - 2\beta }}\\
 = \frac{{2\mu {\varepsilon ^{ - \eta }}}}{{B{{\left( {\bf{X}} \right)}^2}}}\hat{\bm{\rho}}  \times \left[ {{\nabla _{{{\bf{A}}_\alpha }}} \times \left( {\hat{\bm{\rho}}  \times \left( {{\nabla _{{{\bf{A}}_\alpha }}} \times \left( {\left( {\hat{\bm{\rho}} \cdot{\partial _t}{{\bf{A}}_\alpha }} \right){{\widehat {\bf{v}}}_ \bot }} \right)} \right)} \right)} \right]\cdot d{\bf{X}}\\
 = \frac{{2\mu {\varepsilon ^{ - \eta }}}}{{B{{\left( {\bf{X}} \right)}^2}}}\hat{\bm{\rho}} \cdot\left[ {{{\left( {\hat{\bm{\rho}} \cdot\nabla } \right)}^2}{\partial _t}{{\bf{A}}_\alpha }} \right]{\widehat {\bf{v}}_ \bot }
\end{array}
\end{equation}
Therefore, the effective contribution of effect of spatial gradient of $\nabla \phi_\sigma$ is
\begin{equation}\label{167}
\begin{array}{l}
{\varepsilon ^{3 + \eta  - 2\beta }}{\left\langle {{\Gamma _{3 + \eta  - 2\beta }}} \right\rangle _{\bf{X}}} = \frac{1}{{24}}{\varepsilon ^{3 + \eta }}\left\langle {{{\left( {{L_{{\bf{g}}_1^{\bf{X}}}}{L_{{\bf{g}}_1^{\bf{X}}}}{L_{g_\eta ^\mu }}{L_{{\bf{g}}_1^{\bf{X}}}}{\gamma _0}} \right)}_{3 + \eta  - 2\beta }}} \right\rangle \\
 = \frac{1}{{12}}\frac{{{\varepsilon ^3}\mu }}{{B{{\left( {\bf{X}} \right)}^2}}}\left\langle {\hat{\bm{\rho}} \cdot\left[ {{{\left( {\hat{\bm{\rho}} \cdot\nabla } \right)}^2}{\partial _t}{{\bf{A}}_\alpha }} \right]{{\widehat {\bf{v}}}_ \bot }} \right\rangle
\end{array}
\end{equation}

Similar to Eq.(\ref{a152}), it can be derived that
\begin{equation}\label{a159}
{\left( {{L_{{\bf{g}}_1^{\bf{X}}}}{L_{{\bf{g}}_1^{\bf{X}}}}{L_{g_\eta ^\mu }}{L_{{\bf{g}}_1^{\bf{X}}}}{\gamma _0}} \right)_{3 + \eta  - \beta  - \tau }} = 0.
\end{equation}

%

%


%
%

\newpage
\section*{References}


\end{document}